\documentclass[]{sjtu_sai}
\usepackage[toc,page,header]{appendix}
\usepackage{minitoc}
\usepackage{algorithm}
\usepackage{algorithmicx}   
\usepackage{algpseudocode}
\usepackage{booktabs}
\usepackage{multirow}
\usepackage{siunitx}
\usepackage{makecell}
\usepackage{mathtools}
\usepackage{tabularx,array}
\usepackage[table]{xcolor}
\usepackage{graphicx}
\usepackage{wrapfig}
\usepackage{xspace}
\usepackage{amsmath}
\usepackage{amsfonts}

\usepackage[utf8]{inputenc} 
\usepackage[T1]{fontenc}    
\usepackage{hyperref}       
\usepackage{url}            
\usepackage{booktabs}       
\usepackage{amsfonts}       
\usepackage{nicefrac}       
\usepackage{microtype}      
\usepackage{xcolor}         

\usepackage[table]{xcolor}   
\usepackage{pifont}          

\usepackage{amssymb}
\usepackage{mathtools}
\usepackage{amsthm}
\usepackage[most]{tcolorbox}
\usepackage{enumitem}

\usepackage{multirow}

\usepackage{wrapfig}
\usepackage{subcaption}
\theoremstyle{plain}

\theoremstyle{definition}

\theoremstyle{remark}

\title{OmniPresent: Generating Coherent \\ Presentation Suites from Scientific Papers}

\definecolor{PineGreen}{HTML}{007C4A} 
\definecolor{OrangeRed}{HTML}{FF4500} 
\definecolor{mycolor}{gray}{0.92}    

\definecolor{mydarkblue}{rgb}{0,0.08,0.45}
\definecolor{cvprblue}{rgb}{0.21,0.49,0.74}
\definecolor{mydarkblue}{rgb}{0.21,0.49,0.74}

\newcommand{\ourmethod}{OmniPresent}
\newcommand{\ourbench}{OmniPreBench}

\hypersetup{colorlinks=true,citecolor=cvprblue,linkcolor=cvprblue}

\author[1*]{Qianli Ma}
\author[1*]{Jipeng Xiao}
\author[1*]{Siyu Wang}
\author[1]{Zhiheng Tian}
\author[1]{\\ Wangyu Feng}
\author[2]{Shibo Wang}
\author[1]{Chang Guo}
\author[1]{Shuochen Chang}
\author[1]{\\ Qingyang Liu}
\author[1,\dagger]{Zhipeng Zhang}

\makeatletter
\renewcommand\affiliation[2][]{%
  \addtolist[#1]{#2}{\affiliationlist}{\affiliationformat}{\\}}
\makeatother

\affiliation[1]{AutoLab, School of Artificial Intelligence, Shanghai Jiao Tong University}
\affiliation[2]{College of Communication Engineering, Jilin University}
\contribution[*]{Equal Contribution}
\contribution[\dagger]{Corresponding Author}

\abstract{Transforming static research papers into dynamic media such as posters, slides, and videos is essential for effective dissemination but remains a labor-intensive challenge. Existing automated approaches often treat these formats in isolation and consequently fail to maintain semantic consistency across the entire presentation suite. We address this fragmentation by formalizing the task of unified presentation suite generation and proposing \textbf{\ourmethod} to orchestrate the creation of coherent deliverables. Our framework adopts a renderable HTML representation to enable centralized content planning and a self-correcting verify-and-repair loop that actively resolves conflicts across modalities. We further facilitate scalable research in this domain by releasing \textbf{\ourbench}, a comprehensive dataset comprising over one thousand papers with paired artifacts, and establishing a rigorous VLM-based evaluation protocol. Empirical results confirm that our method generates high-quality and faithful presentation suites that significantly surpass strong baselines in both accuracy and visual appeal.
Code and data will be released.}

\checkdata[Project Page]{\url{https://mqleet.github.io/OmniPresent_Page/}}
\begin{document}
\maketitle

\section{Introduction}
\label{sec:intro}


Publishing a paper is often just the first step in the research communication journey. In practice, authors engage with audiences through various additional formats. Posters offer high-density overviews. Slide decks support oral presentations. Videos and interactive project pages provide narrative depth and resource access. However, manually crafting these diverse artifacts is a time-consuming endeavor that places a heavy burden on researchers. Driven by the need for efficiency, recent advancements in large language models have begun to automate these individual formats through systems dedicated to video production~\citep{preacher,paper2video,shi2025presentagent} or slide generation~\citep{evopresent,zheng2025pptagent,liang2025slidegen}. 

However, these existing approaches typically operate in an artifact-centric manner as they treat each format as an independent task~(Fig.~\ref{fig:intro}\textcolor{mydarkblue}{a} and~\ref{fig:intro}\textcolor{mydarkblue}{b}). This fragmentation forces authors to coordinate multiple isolated systems and ultimately fails to address the complex interplay between different media forms. Moreover, such a disjointed approach is particularly risky because maintaining suite-level coherence is inherently difficult under heterogeneous format constraints. For instance, a poster requires spatial condensation while a video demands temporal flow, forcing the same underlying content to be repeatedly compressed and re-expressed. When performed by disjointed systems or manual adaptation, this process frequently leads to inconsistent terminology and mismatched quantitative results. Furthermore, critical nuances such as ablation studies are often inadvertently omitted during these transfers. These discrepancies do more than confuse the audience, as they fundamentally undermine scientific rigor and hinder reproducibility.

\newcommand{\cmark}{\ding{52}}
\begin{figure}[t]
    \centering
    \begin{minipage}[c]{0.45\linewidth} 
        \centering
        \vspace{-2pt}
        \includegraphics[width=\linewidth]{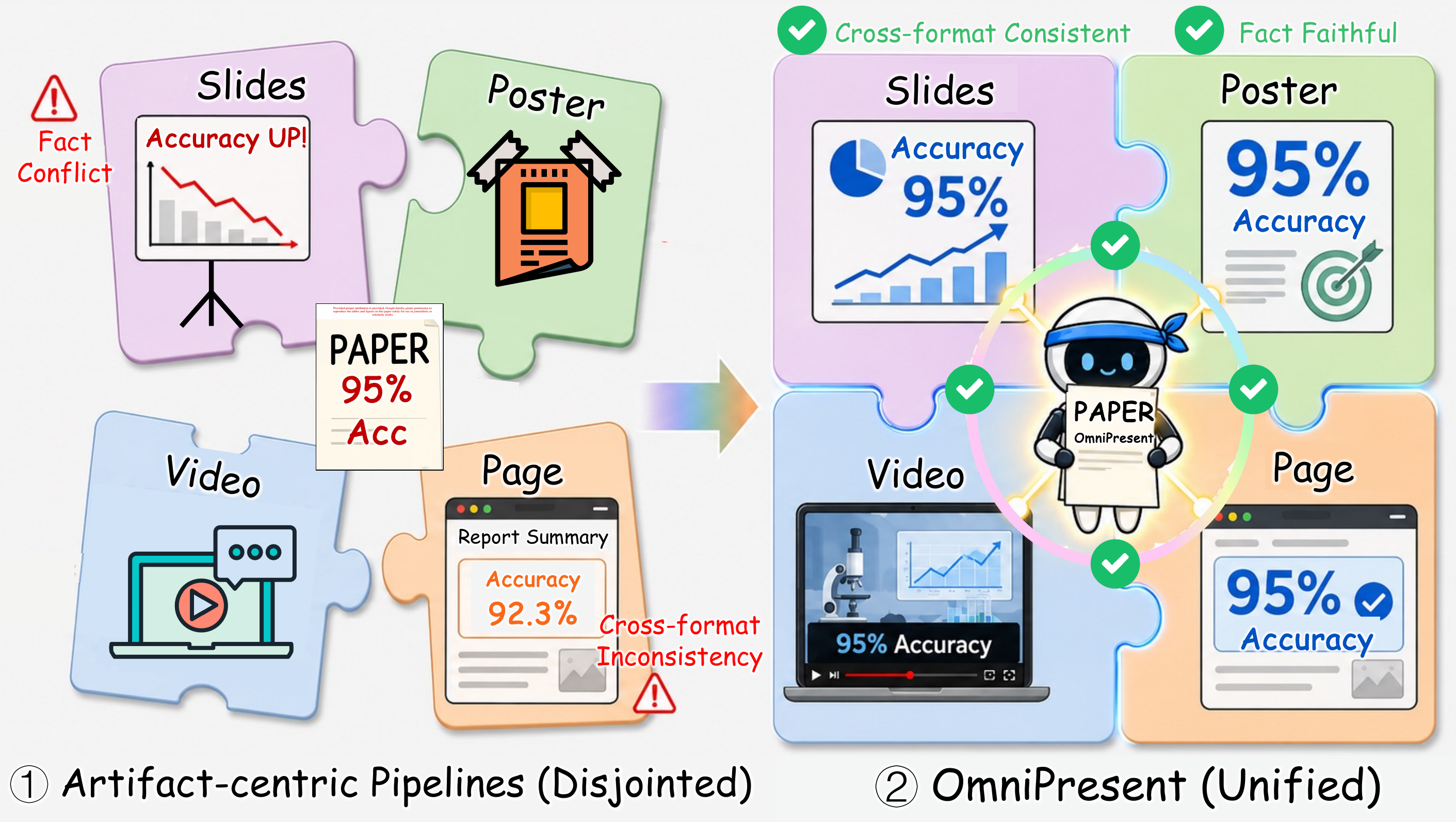}
        \par\vspace{2pt}
        \vspace{-2pt}
        \footnotesize \textbf{(a)} Artifact-centric \emph{v.s.}\ unified suite generation.
    \end{minipage}
    \hfill
    \begin{minipage}[c]{0.52\linewidth} 
        \centering
        \vspace{-2pt}
        \small
        \setlength{\tabcolsep}{3pt} 
        \renewcommand{\arraystretch}{1.15}
        \resizebox{0.8\linewidth}{!}{%
        \begin{tabular}{lcccc}
            \toprule
            \textbf{Method} & \textbf{Poster} & \textbf{Slides} & \textbf{Video} & \textbf{WebPage} \\
            \midrule
            Paper2Poster~\cite{pang2025paper2poster}   & \cmark &            &            &            \\
            PPTAgent~\cite{zheng2025pptagent}       &            & \cmark &            &            \\
            EvoPresent~\cite{ge2025autopresent}    &           & \cmark & \cmark           &            \\
            PresentAgent~\cite{shi2025presentagent}   &            &            & \cmark &            \\
            AutoPage~\cite{autopage}       &            &            &            & \cmark \\
            PaperX~\cite{paperX}       & \cmark           & \cmark            &            & \\
            \cmidrule(lr){1-5}
            \textbf{\ourmethod}             & \cmark & \cmark & \cmark & \cmark \\
            \bottomrule
        \end{tabular}%
        }
        \par\vspace{2pt}
        \footnotesize \textbf{(b)}~Capability coverage across formats.
    \end{minipage}
    \vspace{-5pt}
    \caption{\textbf{From artifact-centric pipelines to unified suite generation.}
\textbf{(a)} Independently generating different artifacts can cause cross-format inconsistencies, such as mismatched claims, numbers, or terminology. OmniPresent unifies generation with cross-artifact checking to improve factual faithfulness and suite-level consistency.
\textbf{(b)} Compared with prior systems that cover only a subset of formats, OmniPresent supports all four formats within one framework.}
    \label{fig:intro}
    \vspace{-10pt}
\end{figure}
Despite the urgency of this challenge, progress is currently limited by the lack of a common evaluation foundation. Specifically, the field lacks a unified framework to assess the generation of a coherent multi-format suite from a single source paper. Without a dataset of paired papers and presentation suites, researchers cannot systematically compare methods or diagnose failure modes regarding cross-format consistency. Consequently, it remains difficult to develop mechanisms that enforce strict alignment in claims and numbers. This absence of scalable evaluation protocols creates a significant barrier to improving the accessibility and faithfulness of automated research dissemination.

To bridge this gap, we formalize the task of \emph{paper to presentation suite generation}. Given a research paper, the goal is to synthesize a comprehensive set of deliverables spanning multiple formats, including a poster, slides, video, and page, as shown in Fig.~\ref{fig:intro}\textcolor{mydarkblue}{a}. This generation process is governed by four essential requirements:
(i) \textbf{faithfulness}, ensuring that salient claims and quantitative results are strictly grounded in evidence from the source paper such as linked sections or figures.
(ii) \textbf{coverage}, capturing the central contributions, methodology, and main experimental findings.
(iii) \textbf{cross-format consistency}, maintaining strict alignment in terminology, core claims, and numbers across all artifacts, as shown in Fig.~\ref{fig:intro}\textcolor{mydarkblue}{a}.
and (iv) \textbf{format compliance}, adhering to the unique structural and communicative constraints of each medium.

To tackle this complex generation task while satisfying the rigorous constraints outlined above, we propose \textbf{\ourmethod}, a unified framework designed to make suite-level coherence enforceable. A key design decision in our framework is the adoption of renderable HTML as the universal underlying representation for all deliverables. We choose HTML because it offers a flexible, structured medium capable of rendering both static layouts (for posters and slides) and dynamic flows (for videos and project pages). Crucially, unlike pixel-based formats, the structured nature of the DOM tree enables precise, code-level verification, directly addressing the need for automated consistency checking. Built upon this foundation, \ourmethod~first performs shared parsing to construct a canonical content plan and evidence map. It then generates each artifact under format constraints while referencing this shared plan, and applies a cross-artifact verify-and-repair loop to detect and resolve conflicts across outputs, such as mismatched numbers, terminology drift, and unsupported claims. 

To support principled and scalable evaluation of this suite-level task, we introduce \textbf{\ourbench}, a benchmark of paired multi-format presentation artifacts curated from recent top-tier AI conference papers. The benchmark comprises 1,037 papers with associated posters, slides, videos, and project pages. To operationalize the assessment, we further propose a scalable LLM-as-judge rubric that evaluates two critical dimensions: content fidelity (via Fact-Score, Content Accuracy, and QA) and presentation quality (via Readability, Visual Aesthetics, and Layout).


Our contributions are summarized as follows: \ding{170} \textbf{Task.} We formalize \emph{unified presentation suite generation}, a task requiring the synthesis of coherent poster, slides, video, and project page from a single paper. \ding{170} \textbf{Method.} We propose \ourmethod, a multi-agent framework that enforces suite-level coherence via centralized content planning and a novel verify-and-repair loop to resolve cross-format conflicts. \ding{170} \textbf{Data\&Bench.} We introduce \ourbench, a large-scale benchmark of paired multi-format data, accompanied by a scalable VLM-based evaluation protocol for quality and consistency. \ding{170} \textbf{Results.} Extensive experiments demonstrate that \ourmethod~produces faithful and high-quality artifacts, significantly outperforming strong baselines.
\section{Related Works}
\label{sec:related}

\noindent\textbf{LLM-based Scientific Agents.} 
Large Language Models (LLMs) have progressed from standalone text generators to \emph{agentic} systems embedded in collaborative frameworks~\citep{wang2024survey,xi2025rise,yao2023react,Chase_LangChain_2022,wu2024autogen}. This shift equips them with the ability to plan~\citep{yao2023react}, invoke tools~\citep{qu2025tool,li2025search}, and carry out multi-step reasoning~\citep{fu2023improving} in pursuit of higher-level goals.
In scientific workflows, such agents show strong potential to automate the research lifecycle, spanning literature surveys~\citep{wang2024autosurvey}, scholarly writing~\citep{weng2025cycleresearcher}, experimental reproducibility~\citep{seo2025paper2code,paper2agent}, and even in peer-review process~\citep{ma2026paper2rebuttal,zhu2025deepreviewimprovingllmbasedpaper}. 
Paper2Agent~\citep{paper2agent} further advances this direction by turning static papers into interactive agents for knowledge retrieval. 
More recently, this agentic paradigm has also been applied to the \emph{dissemination} stage, where agents generate presentation artifacts to communicate completed research across different media formats.



\noindent\textbf{Automated Research Dissemination.}~
Turning a finished paper into compelling dissemination artifacts is essential for research impact. Early template-driven pipelines were brittle and struggled to translate dense scholarly content into coherent narratives~\citep{hu2013ppsgen,paramita2016tailored,sun2021d2s,item1}. Recent progress in LLMs has enabled agentic systems that can both generate content and reason about visual structure, substantially advancing automated presentation generation.
This progress spans multiple formats. For slide decks, PPTAgent~\citep{zheng2025pptagent} and SlideSpawn~\citep{kumar2024slidespawn} distill papers into modular slides, while PreGenie~\citep{pregenie} and AUTOPRESENT~\citep{ge2025autopresent} emphasize narrative structure and visual design; EvoPresent~\citep{evopresent} and Talk to Your Slides~\citep{talktoslides} add refinement through editing. For posters, Paper2Poster~\citep{pang2025paper2poster}, P2P~\citep{sun2025p2p}, and PosterGen~\citep{postergen} adopt layout-aware multi-agent workflows, often guided by VLM feedback. For video, Paper2Video~\citep{paper2video}, Preacher~\citep{preacher}, PresentAgent~\citep{shi2025presentagent}, and AutoPR~\citep{autopr} generate video abstracts and narration for broader outreach. AutoPage~\citep{autopage} builds project webpages as interactive entry points.
Yet, most efforts still advance \emph{one artifact at a time}. In this work, we propose the task of unified presentation suite generation.

\section{\ourmethod}
\label{sec:omnipresent}

\begin{figure*}[t]
            \centering            
            \includegraphics[width=\textwidth]{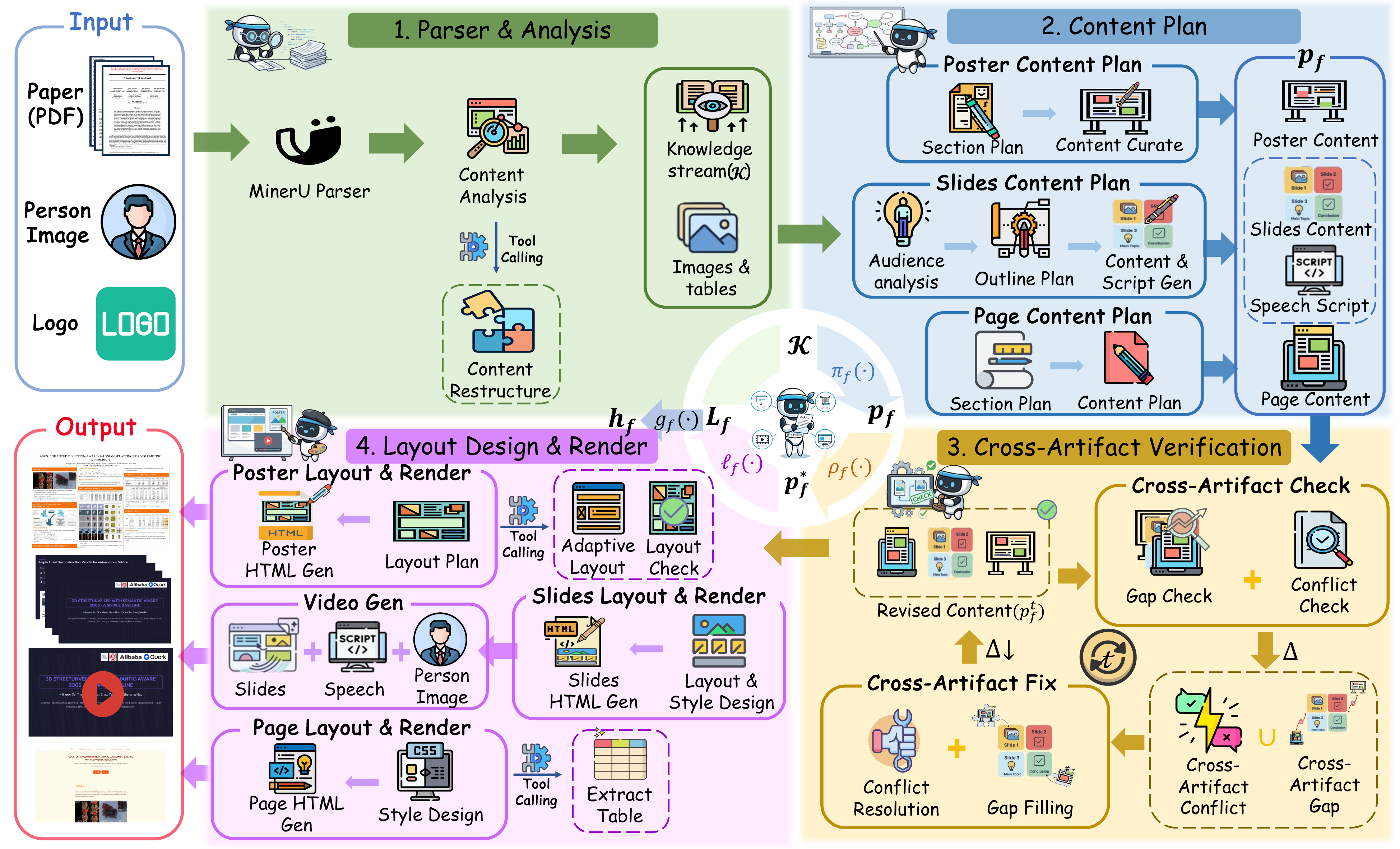}
            \vspace{-15pt}
            \caption{\textbf{Overview of \ourmethod.}~Given a paper PDF, the parser and analysis module extracts structured resources and a shared knowledge stream. Format-specific planners draft content for the project page, poster, and slides, while a Cross-Artifact Verification loop detects and fixes conflicts across outputs. Layout and style agents then render each deliverable as executable HTML.}
            \vspace{-20pt}
     \label{fig:pipeline}
\end{figure*}

We propose \textbf{\ourmethod}, a unified framework that transforms a static scientific paper into a coherent presentation suite. As illustrated in Fig.~\ref{fig:pipeline}, \ourmethod~follows a \textit{Plan-Verify-Render} paradigm.
Unlike traditional frameworks that generate artifacts independently, we introduce a shared \textit{Knowledge Stream} to ground all content and a \textit{Cross-Artifact Verification} loop to ensure suite-level consistency.

\subsection{Problem Formulation and Notation}
\label{subsec:define}

Let $x$ denote the input paper (PDF). Our goal is to generate a suite of presentation artifacts $\mathcal{H} = \{h_f\}_{f \in \mathcal{F}}$ in multiple formats. We define the target format set as: $\mathcal{F}=\{\textsc{Page}, \textsc{Poster}, \textsc{Slides}, \textsc{Video}\}.
$
For each format $f \in \mathcal{F}$, the output is a renderable HTML document $h_f$.
These formats convey different levels of information density. To capture this, we define a containment relation over salient semantic units $\mathcal{I}(\cdot)$:
\begin{equation}
\label{eq:information_unit}
\mathcal{I}(\{\textsc{Poster}, \textsc{Page}\}) \subseteq \mathcal{I}(\textsc{Slides}) \subseteq \mathcal{I}(\textsc{Paper}).
\end{equation}
Eq.~\ref{eq:information_unit} imposes a suite-level constraint: the narrative \textsc{Slides} must structurally cover the highlights of the \textsc{Poster} and \textsc{Page}, motivating our cross-artifact verification.
We formulate the generation process as a four-stage pipeline:

\noindent\textbf{Extraction:} The paper $x$ is parsed into structured resources $\mathcal{R}$ (text, figures, tables) and an evidence index $\mathcal{E}$, which maps atomic claims to supporting evidence spans. These assets are distilled into a shared knowledge stream $\mathcal{K}$.

\noindent\textbf{Planning:} For each format $f$, a policy $\pi_f$ generates an initial content plan $p_f$ based on the shared knowledge~$\mathcal{K}$:
\begin{equation}
\label{eq:plan_policy}
    p_f = \pi_f(\mathcal{K}).
\end{equation}

\noindent\textbf{Verification:} A verifier $\rho_f$ refines the plan to resolve conflicts and ensure the coverage constraints defined in Eq.~\ref{eq:information_unit}, producing a verified plan $p_f^*$:
\begin{equation}
\label{eq:fix_conflict_in_setup}
    p_f^* = \rho_f(\{p_f\}_{f \in \mathcal{F}}, \mathcal{E}).
\end{equation}

\noindent\textbf{Layout \& Rendering:} An adaptive layout module $\ell_f$ translates the verified plan $p_f^*$ into a layout specification $L_f$, which is then rendered by generator $g_f$ into final artifact $h_f$:
\begin{equation}
\label{eq:render}
    L_f = \ell_f(p_f^*), \quad h_f = g_f(L_f).
\end{equation}

This decomposition decouples \textit{semantic planning} ($p_f$) from \textit{visual rendering} ($h_f$), allowing us to rigorously verify content consistency before expensive rendering is incurred.

\subsection{Stage 1: Parse and Extraction}
\label{subsec:parsing}

The foundation of \ourmethod~is the transformation of the PDF data into a queryable \textbf{Knowledge Stream} $\mathcal{K}$, built via a two-step pipeline that first extracts structured raw resources from the PDF and then encodes them into evidence-grounded semantic units.

\noindent\textbf{MinerU Parser \& Content Analysis.}~
We first employ the MinerU parser~\citep{wang2024mineruopensourcesolutionprecise} to extract raw elements: text sections, figures, tables, and references. We denote these resources as $\mathcal{R}$. To support evidence grounding, we decompose content into atomic claims (indivisible factual units) and construct an index $\mathcal{E}$ that maps them to their source spans in $\mathcal{R}$. 
A \textit{Content Analysis Agent} then processes these raw outputs. Rather than directly summarizing raw text, it invokes a \textit{restructure tool} to aggregate disjoint text spans into semantic clusters (\textit{e.g.}, grouping ``Ablation Study" text with its corresponding ``Table 3"). The agent then encodes these clusters into the shared knowledge stream $\mathcal{K}$ (detailed schema in App.~\ref{supp:stream}).

\noindent\textbf{Knowledge Stream ($\mathcal{K}$).}~
Unlike a simple summary, $\mathcal{K}$ encodes the paper's content as elementary units $(c,e,t)$, where $c$ is the textual statement, $e \in \mathcal{E}$ is the evidence pointer, and $t$ is a type tag (\textit{e.g.}, contribution, method, result). This centralized representation ensures that all downstream planners ($\pi_f$) condition on the same source of truth, minimizing cross-format drift.

\subsection{Stage 2: Format-Specific Content Planning}
\label{subsec:planning}

Given the shared knowledge stream $\mathcal{K}$, \ourmethod~synthesizes a format-specific content plan $p_f$. We implement each planner $\pi_f$ as a \textbf{chain of specialized sub-agents}, disentangling high-level structural logic from low-level content generation. As detailed in Figure~\ref{fig:pipeline}, this decomposition addresses distinct modality constraints:

\noindent\textbf{Poster Planner ($\pi_{\textsc{Poster}}$)} optimizes for \textit{spatial information density}. The \textit{Section Planning Agent} first establishes the logical framework by decomposing the paper into distinct thematic sections. Subsequently, the \textit{Content Curation Agent} curates content for each section by identifying and synthesizing the most critical findings into fact-grounded bullet points and selecting the most visually compelling figures to ensure a high-impact and engaging presentation.
    
\noindent\textbf{Slides Planner ($\pi_{\textsc{Slides}}$)} optimizes for \textit{adaptive narrative flow}. Unlike rigid summarization templates, \textit{Audience Analysis Agent} synthesizes a high-level storyline by jointly conditioning on the paper genre and user constraints. This capability allows \ourmethod~to generalize across diverse scientific domains and distinct target audiences. Guided by this narrative arc, the \textit{Outline Planning Agent} sequences the logical beats, which drive the \textit{Content \& Script Gen Agent} to synthesize synchronized visuals and speech commentary strictly aligned with the specific presentation context.

\noindent\textbf{Page Planner ($\pi_{\textsc{Page}}$)} optimizes for \textit{hierarchical engagement}. The \textit{Section Planning Agent} performs adaptive restructuring based on the paper genre, consolidating fragmented chapters into a compact navigation structure. Subsequently, \textit{Content Planning Agent} populates this schema by extracting the essence of technical descriptions and embedding figures/tables as inline visual tokens based on semantic relevance, ensuring a coherent linear reading flow.

Crucially, While these planners operate independently to respect medium constraints, they strictly query the shared $\mathcal{K}$ for information, ensuring that the generated $p_f$ minimizing initial hallucinations across all modalities.

\providecommand{\StageTitle}[1]{\vspace{0.3em}\Statex \textbf{\textcolor{blue!70!black}{\textit{$\triangleright$ #1}}}}
\providecommand{\EqRef}[1]{\hfill \textcolor{gray}{\footnotesize (Eq.~\ref{#1})}}

\begin{figure}[t]
\centering
\resizebox{0.98\linewidth}{!}{
\begin{minipage}{0.98\linewidth}

\hrule height 0.8pt
\vspace{4pt}
\noindent\textbf{Algorithm 1} OmniPresent: Unified Presentation Generation
\vspace{4pt}
\hrule height 0.4pt
\vspace{4pt}

\scriptsize
\begin{algorithmic}[1]
\Require Paper $x$; format set $\mathcal{F}$; max verify-repair iterations $T$
\Ensure Renderable HTML suite $\mathbf{h}=\{h_f\}_{f\in\mathcal{F}}$

\StageTitle{Stage 1: Parse and Extraction}
\State $(\mathcal{R},\mathcal{E}) \leftarrow \mathrm{Parser}(x)$
\State $\mathcal{K} \leftarrow \mathrm{Distill}(\mathcal{R},\mathcal{E})$

\StageTitle{Stage 2: Format-Specific Content Planning}
\For{each $f \in \mathcal{F}$}
    \State $p_f^{(0)} \leftarrow \pi_f(\mathcal{K})$ \EqRef{eq:plan_policy}
\EndFor

\StageTitle{Stage 3: Cross-Artifact Verification}
\For{$t = 0, \ldots, T-1$}
    \State $\mathcal{C}_f^{(t)} \leftarrow \mathrm{ExtractClaims}(p_f^{(t)})$ \textbf{for all} $f \in \mathcal{F}$
    \State $\Delta_{\text{conflict}}^{(t)} \leftarrow \mathrm{ConflictCheck}(\{\mathcal{C}_f^{(t)}\}_{f\in\mathcal{F}}, \mathcal{E})$
    \State $\Delta_{\text{gap}}^{(t)} \leftarrow (\mathcal{C}_{\textsc{Poster}}^{(t)} \cup \mathcal{C}_{\textsc{Page}}^{(t)}) \setminus \mathcal{C}_{\textsc{Slides}}^{(t)}$ \EqRef{eq:gap_define}
    \State $\Delta^{(t)} \leftarrow \Delta_{\text{conflict}}^{(t)} \cup \Delta_{\text{gap}}^{(t)}$
    \If{$\Delta^{(t)} = \varnothing$} \textbf{break} \EndIf
    \For{each $f \in \mathcal{F}$}
        \State $p_f^{(t+1)} \leftarrow \rho_f(p_f^{(t)}, \Delta^{(t)})$ \EqRef{eq:fix_conflict}
    \EndFor
\EndFor

\State $\mathbf{p}^* \leftarrow \{p_f^{(t)}\}_{f\in\mathcal{F}}$

\StageTitle{Stage 4: Layout Design and Render}
\For{each $f \in \mathcal{F}$}
    \State $L_f \leftarrow \ell_f(p_f^*)$ \EqRef{eq:render}
    \State $h_f \leftarrow g_f(L_f)$ \EqRef{eq:render}
\EndFor

\State \textbf{return} $\mathbf{h}=\{h_f\}_{f\in\mathcal{F}}$
\end{algorithmic}

\vspace{4pt}
\hrule height 0.8pt

\end{minipage}
}
\caption{Overview of OmniPresent's unified presentation generation process.}
\label{alg:omnipresent}
\end{figure}

\subsection{Stage 3: Cross-Artifact Verification}
\label{subsec:verify}

This stage is the core contribution of \ourmethod. Generating artifacts in isolation leads to a ``disconnected suite", where the slides might miss key insights from the Poster and Page~(Details in App.~\ref{supp:discussion}). \ourmethod~introduce a \textbf{Cross-Artifact Verification} loop that acts as a \textit{``Chief Editor"}, enforcing consistency and completeness across the suite.

As shown in Fig.~\ref{fig:pipeline}, the verifier $\rho$ takes the set of initial plans $\mathbf{p} = \{p_f\}$ and iterates through a \textit{Check-and-Fix} cycle:

\textbf{1. Cross-Artifact Check (The Critic).}
The \textit{Cross-Artifact Check Agent} inspects the plans for two types of errors:

\noindent\textbf{Conflict Check ($\Delta_{\text{conflict}}$):} It detects hallucinations or contradictions. By cross-referencing plan items against the evidence index $\mathcal{E}$, it flags claims unsupported by the paper or contradicting other formats (\textit{e.g.}, Poster says "95\% accuracy" while Slides say "92\%").

\noindent\textbf{Gap Check ($\Delta_{\text{gap}}$):} It enforces the \textit{Salience Entailment} constraint. If a finding is salient enough to appear in the \textsc{Poster} or \textsc{Page} (high-highlight formats), it must logically appear in the \textsc{Slides} (narrative backbone). Let $\mathcal{C}_f$ denotes the set of semantic claims extracted from plan $p_f$. Formally, we identify the missing coverage:
\begin{equation}
\label{eq:gap_define}
    \Delta_{\text{gap}} = (\mathcal{C}_{\textsc{Poster}} \cup \mathcal{C}_{\textsc{Page}}) \setminus \mathcal{C}_{\textsc{Slides}},
\end{equation}

\textbf{2. Cross-Artifact Fix (The Editor).}
Upon receiving the Conflict \& Gap issues $\Delta$, the \textit{Cross-Artifact Fix Agent} invokes specific tool sets to patch the plans:
\begin{equation}
\label{eq:fix_conflict}
    p_f^{(t+1)} = \rho_f(p_f^{(t)}, \Delta_{\text{conflict}} \cup \Delta_{\text{gap}}).
\end{equation}
The Agent executes \textbf{Conflict Resolution} to rectify unsupported claims and \textbf{Gap Filling} to incorporate \textit{missing coverage} from $\Delta_{\text{gap}}$ into the slides plan $p_\textsc{Slides}$. Crucially, the agent enforces \textbf{modal synchronization}: any structural modification to the slides triggers a synchronous update to the corresponding speech script, ensuring the audio narrative remains strictly aligned with the narrative backbone. This \textit{Check-and-Fix} cycle iterates until convergence ($\Delta^{(t)} = \emptyset$) or a maximum computational budget $T$ is exhausted, yielding the final verified suite $\mathbf{p}^*$.

\subsection{Stage 4: Layout Design and Render}
\label{subsec:rendering}
Stage 4 transforms the verified plan $p_f^*$ into visual artifacts via a visualization layer decoupled from planning. Utilizing Adaptive Layout Module ($\ell_f$) and Adaptive Render Module ($g_f$), this stage accommodates verification-based modifications (Sec.~\ref{subsec:verify}) without breaking rigid templates. $\ell_f$ maps $p_f^*$ to a flexible specification $L_f$ through specialized streams:

\noindent\textbf{Poster Stream.}
The \textit{Layout Plan Agent} organizes section interiors by information density, then uses a Compact Packer tool to optimize visual block arrangement (masonry layout) for minimal area. A Layout Check tool then checks the layout, adjusting fonts or margins to fix overflows.

\noindent\textbf{Slides Stream.}
Managed by the \textit{Layout \& Style Design Agent}, this stream ensures visual hierarchy and consistency. The agent first executes \textbf{Style Design} to unify aesthetics based on $p_f^*$ and audience analysis (Sec.~\ref{subsec:planning}). It then performs \textbf{Layout Design} for each slide, structuring space according to narrative needs to optimize information delivery.

\noindent\textbf{Page Stream.}
The Style Design Agent improves readability and supports personalized styles. It employs the ``Table Extraction'' tool to reconstruct tables into web-compliant formats before styling. Complex layout design is bypassed as HTML naturally adapts to webpage structures.

Finally, $g_f$ generates executable HTML from $L_f$. A Video Gen module then synthesizes rendered slides with speech scripts (from $p_f^*$) and a presenter image into a synchronized audio-visual video.

\section{\ourbench}
\label{sec:bench}

\begin{figure*}[t]
            \centering
            \includegraphics[width=0.9\textwidth]{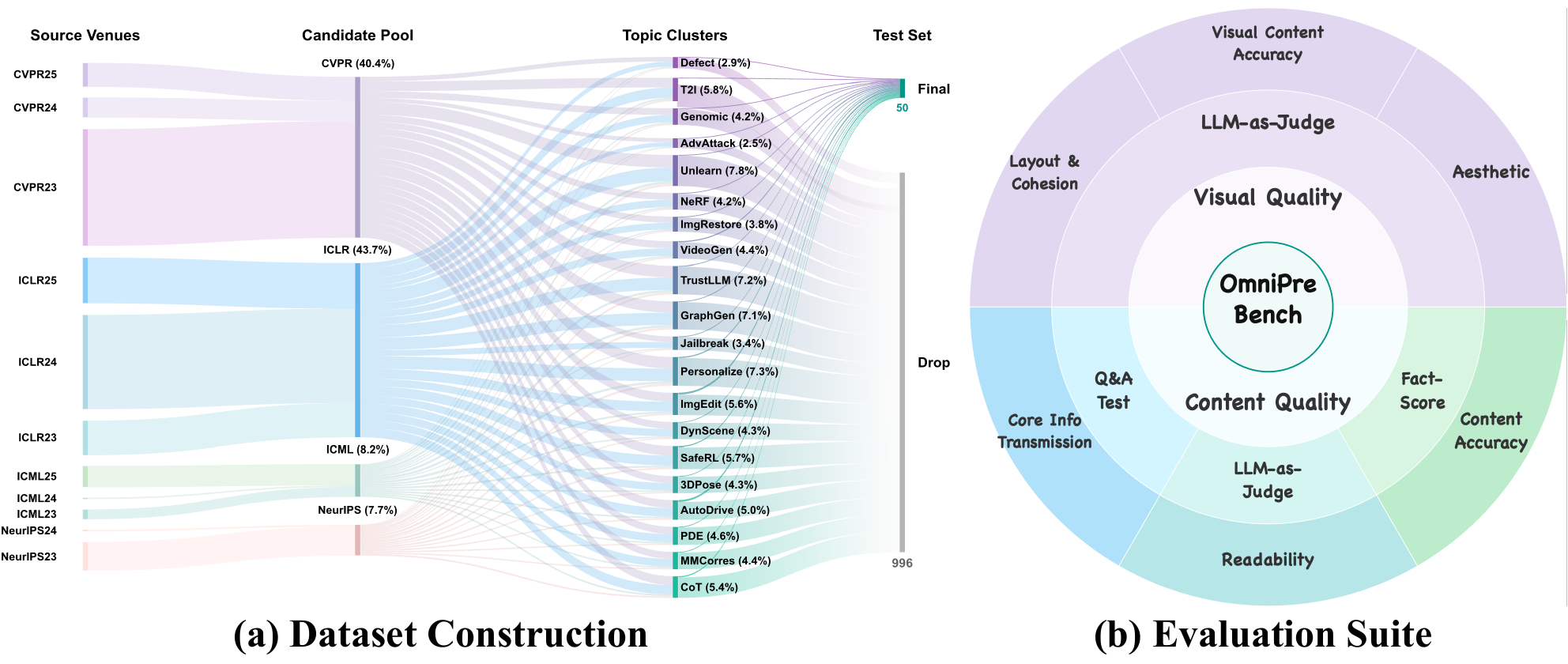}
            \vspace{-10pt}
            \caption{\textbf{Overview of \ourbench.} (\textbf{a}) Dataset construction and testset curation. Papers are collected from multiple venue-year sources, filtered into a qualified candidate pool, clustered by topic, and then stratified to form the final testset. (\textbf{b}) Evaluation suite design. The inner ring defines two top-level dimensions, \emph{Content Quality} and \emph{Visual Quality}. The middle ring shows the evaluation mechanisms.
            The outer ring lists the corresponding fine-grained criteria for each dimension.}
            \vspace{-18pt}
     \label{fig:benchmark}
\end{figure*}
\subsection{Evaluation Dataset}
\label{subsec:data_source}

\noindent\textbf{Data Source.}~
We construct \ourbench~by crawling AI conference papers spanning 2023 to 2025, and retaining only entries with a complete suite of publicly available dissemination artifacts, as illustrated in Fig.~\ref{fig:benchmark}\textcolor{mydarkblue}{a}. Each benchmark sample includes the paper PDF together with its associated project page, poster, slide deck, and presentation video. After filtering by this resource-completeness criterion, we obtain 1{,}046 qualified papers from a curated set of major venues and years, covering a broad range of topics and modalities. Full venue-year coverage, filtering rules, and source statistics are provided in Appendix~\ref{supp:constrct_bench_data}.

\noindent\textbf{Testset Selection.}~
From the qualified pool, we curate a 50-paper testset that prioritizes both diversity and impact. We first compute paper embeddings using SPECTER2 and perform clustering to obtain a semantics-aware partition of topics. We then select papers to ensure broad cluster coverage and venue-year balance, while favoring higher-impact papers measured by citation counts. Detailed sampling procedures and hyperparameters are reported in Appendix~\ref{supp:constrct_bench_data}.

\subsection{Evaluation Metrics}
\label{subsec:evaluation_metric}
We evaluate generated artifacts using a multi-dimensional framework (Fig.~\ref{fig:benchmark}\textcolor{mydarkblue}{b}) with two top-level dimensions, \emph{Content Quality} and \emph{Visual Quality}. We report six metrics, computed with scalable LLM-assisted protocols, including fact extraction and verification, Q\&A-based testing, and rubric-based judging. Details of metrics are reported in Appendix~\ref{supp:constrct_bench_metric}.

\noindent\textbf{Content Quality.}~
We assess factual correctness, information preservation, and writing quality along three axes.
\textbf{\textit{(i)}~Fact-Score}~\citep{min2023factscore} measures whether salient paper facts are preserved in the generated artifact. Using Gemini-3-Flash, we first extract a set of atomic facts $\mathcal{F}_{src}$ from the source paper, and then verify for each $f\in\mathcal{F}_{src}$ whether it is entailed by the generated content $\mathcal{G}$ (see Eq.~\ref{eq:fact-score} in App.~\ref{supp:bench_details})
\textbf{\textit{(ii)}~Q\&A Test} evaluates core information transmission by generating question-answer pairs from the source paper and testing whether the questions can be answered correctly using only the generated artifact as context; we report accuracy over the question set.
\textbf{\textit{(iii)}~Readability} is scored by an LLM-based rubric that assesses clarity, coherence, and fluency.

\noindent\textbf{Visual Quality.}~
We evaluate visual fidelity and presentation quality with VLM-assisted protocols.
\textbf{\textit{(iv)}~Visual Content Accuracy} checks whether visual elements are correctly rendered and semantically aligned with surrounding text, including faithful formula rendering and relevance between figures/diagrams and adjacent descriptions.
\textbf{\textit{(v)}~Layout \& Cohesion} assesses structural organization, such as grouping and hierarchy, balance of spatial distribution, and whether the layout supports a natural reading order.
\textbf{\textit{(vi)}~Aesthetics} rates overall visual appeal, including typography, color harmony, and professional look-and-feel.

\noindent\textbf{Interaction Validity.}~
For project pages, this metric measures the proportion of functional links that correctly navigate to intended destinations. We use an automated browsing agent to simulate and verify success via VLM-based visual feedback~(see App.~\ref{supp:constrct_bench_metric} for implementation details.)

\section{Experiments}
\label{sec:exp}

We evaluate \ourmethod~on \ourbench~with a focus on suite-level quality, including evidence-grounded faithfulness, coverage of salient contributions, cross-format consistency, and format compliance.
We compare against two competitive baseline families: \emph{(i) strong closed-source models} that generate all artifacts end to end in a single pass, and \emph{(ii) format-specialized agentic systems} designed for individual artifacts. In addition to overall performance, we conduct controlled comparisons that isolate the effect of unified parsing and planning, as well as the cross-artifact verify-and-repair loop.
To move beyond aggregate metrics, we also conduct qualitative case studies to visualize the suite-level advantages of \ourmethod. Due to the content limitation, detailed analyses are provided in App.~\ref{supp:cases}.

\subsection{Experimental Setup}
\label{subsec:experiment_setup}

\noindent\textbf{Baselines.}~
We compare \ourmethod~against two categories of baselines.
First, we evaluate strong end-to-end generation using closed-source models that directly produce all artifacts in a single pass, without explicit suite-level planning or verification. We include Gemini3-Flash~\citep{team2023gemini}, GPT-5.2-mini~\citep{openai2025gpt5} and Grok4-Fast~\citep{xGrok} as representative high-capability proprietary models.
Second, we compare against existing format-specialized agentic systems that are designed for individual artifacts rather than a unified suite, including Paper2Poster~\citep{pang2025paper2poster} for posters, EvoPresent~\citep{evopresent} for slide decks and videos, and AutoPage~\citep{autopage} for project pages.
These methods provide strong, domain-specific pipelines tailored to each format.
We further include PaperX~\cite{paperX}, a concurrent unified framework that uses a Scholar DAG as the intermediate representation to generate multi-format artifacts via adaptive graph traversal.
Additionally, we include \textit{Human-authored} artifacts as a competitive reference to evaluate the gap between automated generation and human standards.

\noindent\textbf{Implementation details.}~
To ensure fair comparisons, we control for the underlying foundation model whenever possible. When comparing with proprietary end-to-end LLM baselines, we instantiate \ourmethod~using the \emph{same} proprietary model as its backbone, so that gains can be attributed to our unified parsing, planning, and cross-artifact verification rather than model capacity.
When comparing with prior format-specialized systems, we also align the backbone model used in their implementations with the one used in \ourmethod, ensuring that improvements primarily reflect differences in system design and suite-level verification rather than differences in base model choice.


\providecolor{paBlue}{RGB}{232,240,254}
\providecolor{paGreen}{RGB}{236,253,245}
\providecolor{paYellow}{RGB}{255,251,235}
\providecolor{paPink}{RGB}{255,240,246}

\newcommand{\paSeprule}{%
  \noalign{\global\arrayrulewidth=0.9pt}\hline
  \noalign{\global\arrayrulewidth=0.4pt}%
}

\newcommand{\paTightSpace}{\addlinespace[2pt]}
\newcommand{\paLooseSpace}{\addlinespace[4pt]}
\newcommand{\mindent}{\hspace{0.8em}}

\newcommand{\SectionRow}[1]{%
  \paSeprule
  \multicolumn{9}{@{}l@{}}{\textbf{\textit{#1}}}\\[-2pt]
}

\newcolumntype{V}{@{\hspace{6pt}\vrule\hspace{6pt}}}

\newcommand{\inc}[1]{\,\textcolor{green!60!black}{\scriptsize{(+#1)}}}

\newcommand{\ValW}{2.8em}
\newcommand{\IncW}{3.4em}
\newcommand{\entry}[2]{%
  \makebox[\ValW][r]{#1}\hspace{1pt}\makebox[\IncW][l]{#2}%
}

\newcommand{\CompW}{4.0em}
\newcommand{\comp}[1]{\makebox[\CompW][r]{#1}}

\begin{table*}[t]
\centering
\caption{\textbf{Evaluation results across our full suite of OmniPreBench.} We report content quality (FactScore, Readability, QA), visual quality (visual content accuracy, aesthetics, layout), interaction success rate (Page only), and compression rate (higher indicates less redundancy). For each proprietary backbone, we compare end-to-end generation with \ourmethod~instantiated using the same backbone; values in parentheses denote the absolute gains of \ourmethod~over its matched end-to-end baseline.}
\label{tab:combined_slides_poster_page_results}
\vspace{-6pt}
\small
\setlength{\tabcolsep}{5pt}
\renewcommand{\arraystretch}{1.12}
\resizebox{\textwidth}{!}{%
\begin{tabular}{lVcccVcccVcVc}
\toprule
\multirow{2}{*}{\textbf{Method}} &
\multicolumn{3}{cV}{\textbf{Content}} &
\multicolumn{3}{cV}{\textbf{Visual}} &
\textbf{Interaction} &
\textbf{Compression} \\
\cmidrule(lr){2-4}\cmidrule(lr){5-7}\cmidrule(lr){8-8}\cmidrule(lr){9-9}
& \shortstack[c]{FactScore$\uparrow$}
& \shortstack[c]{Readability$\uparrow$}
& \shortstack[c]{QA$\uparrow$}
& \shortstack[c]{Content acc$\uparrow$}
& \shortstack[c]{Aesthetics$\uparrow$}
& \shortstack[c]{Layout$\uparrow$}
& \shortstack[c]{Success$\uparrow$}
& \shortstack[c]{Compression Rate} \\

\SectionRow{Slides}
\paTightSpace
\mindent Human-authored &
\entry{0.716}{} & \entry{3.26}{} & \entry{0.711}{} &
\entry{2.22}{}  & \entry{3.53}{} & \entry{3.10}{} &
- & \comp{12.62} \\
\mindent EvoPresent-Gemini-3-flash~\citep{evopresent} &
\entry{0.666}{} & \entry{3.38}{} & \entry{0.713}{} &
\entry{1.74}{}  & \entry{3.38}{} & \entry{2.60}{} &
- & \comp{24.96} \\

\mindent PaperX-Gemini-3-flash~\citep{paperX} &
\entry{0.652}{} & \entry{3.39}{} & \entry{0.725}{} &
\entry{2.63}{}  & \entry{3.98}{} & \entry{3.09}{} &
- & \comp{9.35} \\

\mindent GPT-5-mini &
\entry{0.751}{} & \entry{3.00}{} & \entry{0.716}{} &
\entry{2.27}{}  & \entry{3.00}{} & \entry{2.85}{} &
- & \comp{6.92} \\
\rowcolor{paBlue}
\mindent OmniPresent-GPT-5-mini &
\entry{\textbf{0.830}}{\inc{0.079}} &
\entry{\textbf{3.20}}{\inc{0.20}} &
\entry{\textbf{0.724}}{\inc{0.008}} &
\entry{\textbf{2.84}}{\inc{0.57}} &
\entry{\textbf{3.89}}{\inc{0.89}} &
\entry{\textbf{3.49}}{\inc{0.64}} &
- & \comp{6.74} \\
\paTightSpace
\mindent Gemini-3-flash &
\entry{0.585}{} & \entry{3.18}{} & \entry{0.707}{} &
\entry{1.94}{}  & \entry{2.39}{} & \entry{2.33}{} &
- & \comp{15.18} \\
\rowcolor{paGreen}
\mindent OmniPresent-Gemini-3-flash &
\entry{\textbf{0.816}}{\inc{0.231}} &
\entry{\textbf{3.40}}{\inc{0.22}} &
\entry{\textbf{0.751}}{\inc{0.044}} &
\entry{\textbf{2.75}}{\inc{0.81}} &
\entry{\textbf{4.53}}{\inc{2.14}} &
\entry{\textbf{3.54}}{\inc{1.21}} &
- & \comp{8.55} \\
\paTightSpace

\mindent Grok-4.1-fast &
\entry{0.691}{} & \entry{2.98}{} & \entry{0.705}{} &
\entry{2.04}{}  & \entry{2.00}{} & \entry{2.29}{} &
- & \comp{11.95} \\
\rowcolor{paYellow}
\mindent OmniPresent-Grok-4.1-fast &
\entry{\textbf{0.827}}{\inc{0.136}} &
\entry{\textbf{3.08}}{\inc{0.10}} &
\entry{\textbf{0.723}}{\inc{0.018}} &
\entry{\textbf{2.12}}{\inc{0.08}} &
\entry{\textbf{3.08}}{\inc{1.08}} &
\entry{\textbf{2.33}}{\inc{0.04}} &
- & \comp{8.55} \\

\SectionRow{Poster}
\paTightSpace
\mindent Human-authored &
\entry{0.733}{} & \entry{3.43}{} & \entry{0.710}{} &
\entry{3.05}{}  & \entry{3.50}{} & \entry{3.02}{} &
- & \comp{15.34} \\
\mindent Paper2Poster-Gemini-3-flash~\citep{pang2025paper2poster} &
\entry{0.470}{} & \entry{3.50}{} & \entry{0.659}{} &
\entry{2.88}{}  & \entry{2.05}{} & \entry{3.00}{} &
- & \comp{5.33} \\

\mindent PaperX-Gemini-3-flash~\citep{paperX} &
\entry{0.526}{} & \entry{3.51}{} & \entry{0.664}{} &
\entry{2.03}{}  & \entry{3.02}{} & \entry{2.68}{} &
- & \comp{23.62} \\

\mindent GPT-5-mini &
\entry{0.766}{} & \entry{3.00}{} & \entry{0.730}{} &
\entry{2.85}{}  & \entry{2.27}{} & \entry{2.85}{} &
- & \comp{5.35} \\
\rowcolor{paBlue}
\mindent OmniPresent-GPT-5-mini &
\entry{\textbf{0.779}}{\inc{0.013}} &
\entry{\textbf{3.46}}{\inc{0.46}} &
\entry{\textbf{0.746}}{\inc{0.016}} &
\entry{\textbf{3.26}}{\inc{0.41}} &
\entry{\textbf{3.79}}{\inc{1.52}} &
\entry{\textbf{2.96}}{\inc{0.11}} &
- & \comp{3.30} \\
\paTightSpace
\mindent Gemini-3-flash &
\entry{0.585}{} & \entry{3.44}{} & \entry{0.695}{} &
\entry{2.43}{}  & \entry{3.63}{} & \entry{3.02}{} &
- & \comp{20.29} \\
\rowcolor{paGreen}
\mindent OmniPresent-Gemini-3-flash &
\entry{\textbf{0.755}}{\inc{0.170}} &
\entry{\textbf{3.60}}{\inc{0.16}} &
\entry{\textbf{0.716}}{\inc{0.021}} &
\entry{\textbf{3.16}}{\inc{0.73}} &
\entry{\textbf{3.76}}{\inc{0.13}} &
\entry{\textbf{3.20}}{\inc{0.18}} &
- & \comp{3.75} \\
\paTightSpace
\mindent Grok-4.1-fast &
\entry{0.702}{} & \entry{3.18}{} & \entry{0.713}{} &
\entry{2.06}{}  & \entry{2.59}{} & \entry{2.55}{} &
- & \comp{12.82} \\
\rowcolor{paYellow}
\mindent OmniPresent-Grok-4.1-fast &
\entry{\textbf{0.754}}{\inc{0.052}} &
\entry{\textbf{3.22}}{\inc{0.04}} &
\entry{\textbf{0.717}}{\inc{0.004}} &
\entry{\textbf{3.10}}{\inc{1.04}} &
\entry{\textbf{3.73}}{\inc{1.14}} &
\entry{\textbf{3.04}}{\inc{0.49}} &
- & \comp{3.21} \\

\SectionRow{Page}
\paTightSpace
\mindent Human-authored &
\entry{0.721}{} & \entry{3.12}{} & \entry{0.706}{} &
\entry{3.40}{}  & \entry{3.20}{} & \entry{3.46}{} &
\entry{1.00}{} & \comp{10.89} \\
\mindent AutoPage-Gemini-3-flash~\citep{autopage} &
\entry{0.705}{} & \entry{3.10}{} & \entry{0.701}{} &
\entry{2.90}{}  & \entry{2.07}{} & \entry{3.35}{} &
\entry{0.66}{} & \comp{3.28} \\
\mindent GPT-5-mini &
\entry{0.741}{} & \entry{3.06}{} & \entry{0.710}{} &
\entry{1.82}{}  & \entry{3.25}{} & \entry{3.56}{} &
\entry{0.04}{} & \comp{4.78} \\
\rowcolor{paBlue}
\mindent OmniPresent-GPT-5-mini &
\entry{\textbf{0.801}}{\inc{0.060}} &
\entry{\textbf{4.48}}{\inc{1.42}} &
\entry{\textbf{0.714}}{\inc{0.004}} &
\entry{\textbf{3.28}}{\inc{1.46}} &
\entry{\textbf{3.45}}{\inc{0.20}} &
\entry{\textbf{3.66}}{\inc{0.10}} &
\entry{0.76}{\inc{0.72}} & \comp{8.84} \\
\paTightSpace
\mindent Gemini-3-flash &
\entry{0.485}{} & \entry{3.08}{} & \entry{0.699}{} &
\entry{1.14}{}  & \entry{3.46}{} & \entry{3.02}{} &
\entry{0.02}{} & \comp{13.13} \\
\rowcolor{paGreen}
\mindent OmniPresent-Gemini-3-flash &
\entry{\textbf{0.782}}{\inc{0.297}} &
\entry{\textbf{3.62}}{\inc{0.54}} &
\entry{\textbf{0.712}}{\inc{0.013}} &
\entry{\textbf{3.12}}{\inc{1.98}} &
\entry{\textbf{3.59}}{\inc{0.13}} &
\entry{\textbf{3.96}}{\inc{0.94}} &
\entry{0.92}{\inc{0.90}} & \comp{6.98} \\
\paTightSpace
\mindent Grok-4.1-fast &
\entry{0.666}{} & \entry{2.78}{} & \entry{0.703}{} &
\entry{1.40}{}  & \entry{3.22}{} & \entry{3.60}{} &
\entry{0.04}{} & \comp{9.60} \\
\rowcolor{paYellow}
\mindent OmniPresent-Grok-4.1-fast &
\entry{\textbf{0.791}}{\inc{0.125}} &
\entry{\textbf{3.66}}{\inc{0.88}} &
\entry{\textbf{0.747}}{\inc{0.044}} &
\entry{\textbf{3.42}}{\inc{2.02}} &
\entry{\textbf{3.63}}{\inc{0.41}} &
\entry{\textbf{3.94}}{\inc{0.34}} &
\entry{0.72}{\inc{0.68}} & \comp{12.38} \\

\bottomrule
\end{tabular}%
}
\end{table*}

\subsection{Main Results}
\label{subsec:main_result}

Tab.~\ref{tab:combined_slides_poster_page_results} reports overall results on slides, posters, and project pages under our content and visual metrics.
Overall, the gains are systematic rather than incidental: \ourmethod~improves both content faithfulness and visual presentation across formats while holding the backbone model fixed, indicating that the improvements come from the \textit{suite-level \ourmethod~pipeline} rather than model capacity.

\noindent\textbf{\textit{Findings 1}: \ourmethod~outperforms strong specialized and partially unified baselines without sacrificing per-format quality.}~
Compared with format-specialized systems and the concurrent partially unified baseline PaperX~(only poster and slides), \ourmethod~achieves stronger per-format performance under the same Gemini-3-Flash backbone~(Tab.~\ref{tab:combined_slides_poster_page_results}). 
In slide generation, \ourmethod~improves over EvoPresent and PaperX in \textit{FactScore} (0.816 \textit{v.s.} 0.666/0.652), \textit{QA Accuracy} (0.751 \textit{v.s.} 0.713/0.725), and \textit{Aesthetics} (4.53 \textit{v.s.} 3.38/3.98). 
In poster generation, it also surpasses Paper2Poster and PaperX, \emph{e.g.}, in \textit{FactScore} (0.755 \textit{v.s.} 0.470/0.526) and \textit{Visual Content Accuracy} (3.16 \textit{v.s.} 2.88/2.03). 
This supports our core hypothesis in Sec.~\ref{sec:intro} that sharing a centralized \textit{Knowledge Stream}~(Sec.~\ref{subsec:planning}) and performing \textit{Cross-Artifact Verification}~(Sec.~\ref{subsec:verify}) does not just enforce coherence but actually improves individual artifact quality by grounding them in a unified evidence index.

\noindent\textbf{\textit{Findings 2}: \ourmethod~ensures superior functional integrity and navigational reliability.}~
Beyond static visual quality, we evaluate the interactive functionality of project pages.
As shown in Tab.~\ref{tab:combined_slides_poster_page_results}, \ourmethod~achieves a significantly higher \textit{Interaction Success} rate compared to baselines. 
This success is a direct result of \textit{Evidence Indexing~(Sec.~\ref{subsec:parsing}):} the former ensures that every external link (\textit{e.g.}, GitHub repository) is grounded in the source paper's metadata, while the latter precisely maps these resources to the correct HTML code during the rendering process.

\noindent\textbf{\textit{Findings 3}: Synchronous content-script planning enhances video faithfulness and information accessibility.}~
Tab.~\ref{tab:video_results} compares \ourmethod~against EvoPresent in the video task. 
Even when using the same backbone, our method consistently outperforms the baseline, particularly in \textit{FactScore} and \textit{QA Accuracy}, indicating that our unified knowledge stream provides more reliable grounding for temporal content in video. 
Furthermore, an inter-table analysis between Tab.~\ref{tab:combined_slides_poster_page_results}~(see \textit{slides}) and Tab.~\ref{tab:video_results} (video) reveals an interesting \textit{``readability boost"}: the Readability score for videos is consistently higher than that of the corresponding slides.
While standard sequential pipelines derive narration only after slide content is fixed, \ourmethod~generates \textit{speaker notes} in tandem with the slide content planning phase.
This integrated approach, as designed in our \textit{Content Planning Agent} in Sec.~\ref{subsec:planning}, ensures a higher degree of alignment between visual bullet points and verbal explanations, which directly translates to improved readability. 

\providecolor{paBlue}{RGB}{232,240,254}
\providecolor{paGreen}{RGB}{236,253,245}
\providecolor{paYellow}{RGB}{255,251,235}
\providecolor{paPink}{RGB}{255,240,246}

\providecommand{\paSeprule}{%
  \noalign{\global\arrayrulewidth=0.9pt}\hline
  \noalign{\global\arrayrulewidth=0.4pt}%
}

\providecommand{\inc}[1]{\,\textcolor{green!60!black}{\scriptsize{(+#1)}}}
\providecommand{\dec}[1]{\,\textcolor{red!70!black}{\scriptsize{(#1)}}}
\providecommand{\neu}[1]{\,\textcolor{black!45}{\scriptsize{(#1)}}}

\providecommand{\ValWv}{2.3em}
\providecommand{\IncWv}{2.2em}
\providecommand{\entryv}[2]{%
  \makebox[\ValWv][r]{#1}\hspace{1pt}\makebox[\IncWv][l]{#2}%
}

\providecommand{\ValW}{2.8em}
\providecommand{\IncW}{3.4em}
\providecommand{\entry}[2]{%
  \makebox[\ValW][r]{#1}\hspace{1pt}\makebox[\IncW][l]{#2}%
}

\begin{table*}[t] 
\centering

\begin{minipage}[t]{0.54\textwidth}
\centering
\caption{\textbf{Evaluation results of video.} We report content metrics (FactScore, Readability, QA) and compression for video artifacts. For each backbone, the improvements of \ourmethod~in parentheses are computed relative to EvoPresent.}
\vspace{-5pt}
\label{tab:video_results}

\small
\setlength{\tabcolsep}{3pt}
\renewcommand{\arraystretch}{1.12}

\resizebox{\linewidth}{!}{%
\begin{tabular}{lccc}
\toprule
\textbf{Method} &
\textbf{FactScore$\uparrow$} &
\textbf{Readability$\uparrow$} &
\textbf{QA$\uparrow$} \\
\midrule
Human-authored &
\entryv{0.849}{} & \entryv{3.51}{} & \entryv{0.759}{}  \\
EvoPresent-Gemini-3-flash &
\entryv{0.687}{} & \entryv{3.50}{} & \entryv{0.732}{}  \\
OmniPresent-Gemini-3-flash &
\entryv{0.865}{\inc{0.179}} & \entryv{3.53}{\inc{0.03}} & \entryv{0.773}{\inc{0.041}}\\
OmniPresent-GPT-5-mini &
\entryv{0.856}{\inc{0.170}} & \entryv{3.39}{\dec{-0.11}} & \entryv{0.757}{\inc{0.025}} \\
OmniPresent-Grok-4.1-fast &
\entryv{0.813}{\inc{0.126}} & \entryv{3.24}{\dec{-0.26}} & \entryv{0.746}{\inc{0.014}} \\
\bottomrule
\end{tabular}%
}
\end{minipage}%
\hfill
\begin{minipage}[t]{0.44\textwidth}
\centering
\caption{\textbf{Ablation study on CV.} We ablate the CV on \ourmethod~and report the absolute change w.r.t.\ w/o CV in parentheses.}
\label{tab:ablation_cross_valid}
\vspace{-3pt} 

\small
\setlength{\tabcolsep}{4pt}
\renewcommand{\arraystretch}{1.2}

\resizebox{\linewidth}{!}{%
\begin{tabular}{@{}llcccc@{}}
\toprule
\multirow{2}{*}{\textbf{Setting}} & \multirow{2}{*}{\textbf{Method}} & \multicolumn{3}{c}{\textbf{Content Quality}} & \textbf{Compression} \\
\cmidrule(lr){3-5} \cmidrule(l){6-6}
 & & FactScore$\uparrow$ & Readability$\uparrow$ & QA$\uparrow$ & Rate \\
\midrule

\multirow{2}{*}{\textbf{Page}}
 & w/o CV  & \entry{0.782}{}            & \entry{3.44}{}            & \entry{0.698}{}            & 3.67 \\
 & w/ ~CV  & \entry{0.810}{\inc{0.028}} & \entry{3.42}{\dec{-0.02}} & \entry{0.712}{\inc{0.014}} & 3.75 \\
\midrule

\multirow{2}{*}{\textbf{Slides}}
 & w/o CV  & \entry{0.774}{}            & \entry{3.48}{}             & \entry{0.718}{}            & 6.58 \\
 & w/ ~CV  & \entry{0.816}{\inc{0.042}} & \entry{3.48}{\neu{+0.00}}  & \entry{0.750}{\inc{0.032}} & 6.98 \\
\midrule

\multirow{2}{*}{\textbf{Poster}}
 & w/o CV  & \entry{0.732}{}            & \entry{3.58}{}            & \entry{0.700}{}            & 8.31 \\
 & w/ ~CV  & \entry{0.755}{\inc{0.023}} & \entry{3.60}{\inc{0.02}}  & \entry{0.716}{\inc{0.016}} & 8.55 \\
\midrule

\multirow{2}{*}{\textbf{Video}}
 & w/o CV  & \entry{0.793}{}            & \entry{3.52}{}            & \entry{0.732}{}            & 6.81 \\
 & w/ ~CV  & \entry{0.865}{\inc{0.072}} & \entry{3.53}{\inc{0.01}}  & \entry{0.773}{\inc{0.041}} & 6.03 \\

\bottomrule
\end{tabular}%
}
\end{minipage}

\vspace{-20pt}
\end{table*}

\subsection{Human Preference Study}
\label{subsec:user_study}
\noindent\textbf{Study design.}~
We conduct a user study to (i) validate that \ourmethod~improves human-perceived visual quality over competitive baselines and (ii) assess the alignment between our VLM-based visual rubric and human judgment.
Participants evaluate generated suites on a 5-point Likert scale across three visual quality dimensions. 
By comparing these human scores with our automated metrics, we evaluate the reliability of our evaluation framework. Detailed settings are provided in Appendix~\ref{supp:user_guidance}.

\noindent\textbf{\textit{Findings 4}: Human preference supports both method effectiveness and rubric validity.}~
Experimental results in Tab.~\ref{tab:user_study_visual_sc} indicate a strong human preference for \ourmethod~over both end-to-end baselines and specialized systems. 
Participants consistently award \ourmethod~the highest mean ratings in \textit{Visual Aesthetics} and \textit{Layout} across all formats, underscoring its capability to produce professional-grade presentation materials. 
Crucially, we observe a high correlation between human ratings and our VLM-based scores; the preference ordering produced by humans perfectly mirrors the rankings generated by our automated rubric. 
This alignment suggests that our \ourbench~effectively captures the nuanced visual attributes valued by users.

\subsection{Ablation Study}
\label{subsec:ablation}

\noindent\textbf{Ablation setting.}~
To isolate the effect of our \textit{Cross-Artifact Verification}~(CV, Sec.~\ref{subsec:verify}), we conduct a controlled ablation by removing the CV module while keeping all other components, prompts, and decoding settings identical. We evaluate both configurations using Gemini-3-Flash.
As shown in Tab.~\ref{tab:ablation_cross_valid}, we assess per-format performance across webpages, slides, and posters, focusing on \textit{Content Quality} and \textit{Compression} (Com), the latter of which measures the efficiency of information density.

\noindent\textbf{\textit{Findings 5}: CV is critical for suite-level faithfulness and information-dense research suite.}~
As reported in Tab.~\ref{tab:ablation_cross_valid}, removing the CV leads to a consistent decline in both \emph{FactScore} and \emph{QA Accuracy}. 
\textit{This result directly validates the ``Verify-and-Repair" loop} described in Sec.~\ref{subsec:verify}. By cross-referencing generated claims against the \textit{Evidence Index} in Sec.~\ref{subsec:parsing}, the CV identifies and rectifies hallucination drifts that often occur during independent format planning. 
The boost in QA scores further indicates that CV ensures higher information coverage, as the \textit{Chief Editor} (see Sec.~\ref{subsec:verify}) proactively checks for missing salient contributions across the suite.
Notably, while \emph{Readability} remains stable, which suggests that the backbone's linguistic fluency is unaffected, the \emph{Com} metric improves with CV. 
This confirms that our verification process \textit{not only filters inaccuracies but also prunes redundant or repetitive content, leading to a more concise and professional suite}. 

\section{Conclusion}
In this work, we introduce \textit{\ourmethod}, a unified system that generates a coherent presentation suite in executable HTML, including a project page, poster, slides, and video, from a single scientific paper.
By leveraging shared parsing and content planning, \ourmethod~ensures consistency across formats while grounding each artifact in the original paper.
We further introduced a cross-artifact verification loop that detects and resolves conflicts, improving fidelity and reducing cross-format drift in artifact generation.
To support principled, large-scale evaluation of this suite-level task, we build a new benchmark, \textit{\ourbench}, by collecting paired multi-format artifacts for recent AI conference papers, enabling systematic assessment of faithfulness, coverage, cross-format consistency, and format compliance.
Experiments show that \ourmethod~consistently improves groundedness and suite-level consistency over strong independent-generation baselines. We hope this work will support future progress on reliable, scalable scientific communication.

\newpage
\appendix
\onecolumn



\section{Details of \ourbench}
\label{supp:bench_details}

\subsection{Construction of the Dataset}
\label{supp:constrct_bench_data}
\noindent\textbf{Data Sources.}~
We construct our benchmark by initially collecting a broad set of 31,521 candidate papers from 13 major AI conference~(\textit{e.g.}, AAAI, ACL, CVPR, NeurIPS, ICLR, ICML) tracks spanning 2023 to 2025, stored in our raw data repository.
However, to enable a comprehensive multimodal evaluation, we enforce a strict resource completeness criterion: valid benchmark entries must feature not only the full paper PDF but also a suite of ground-truth multimedia assets, including a project webpage, a poster, a slide deck, and a presentation video.

Due to the varying availability of these open-access resources across different venues, we narrow our focus to a curated subset of 11 specific venue-year combinations: \emph{CVPR} (2023-2025), \emph{ICLR} (2023-2025), \emph{ICML} (2023-2025), and \emph{NeurIPS} (2023-2024). 
Unlike other venues such as AAAI or ACL, which were excluded from the final selection due to sparse multimedia hosting, these chosen conferences consistently utilize platforms (\textit{e.g.}, YouTube, OpenReview, SlidesLive) that archive high-quality supplementary materials. 
This filtering process yields a high-quality candidate pool of 1,037 papers, ensuring that every sample in our benchmark possesses the necessary ground-truth data for cross-verifying generated content against multiple modalities.

\noindent\textbf{Data Analysis.}~The selected candidate pool represents a diverse cross-section of the AI landscape, capturing distinct shifts in community focus over the past three years. 
In 2023, the field was dominated by foundational breakthroughs in \textit{Reasoning} (\textit{e.g.}, Chain-of-Thought, Cluster 20) and \textit{Personalized Image Synthesis} (\textit{e.g.}, DreamBooth, Cluster 12). 
This focus expanded in 2024 to address \textit{Multimodal Safety \& Robustness} (\textit{e.g.}, Jailbreak Attacks, Cluster 11) and specialized scientific applications (\textit{e.g.}, Genomics, Cluster 3). 
Most recently, in 2025, we observe a significant pivot towards \textit{Temporal \& Embodied Intelligence}, characterized by a surge in Video Generation and Robotic Manipulation research. 
This thematic progression ensures that agents are challenged to synthesize and visualize diverse technical narratives, ranging from abstract theoretical frameworks to dynamic visual demonstrations into coherent multimodal artifacts.

\noindent\textbf{Testset Filtering.}~To establish a robust and representative benchmark, we curate a core test set of 50 papers from this qualified pool of 1,037 works. 
We employ a \textit{semantics-aware stratified sampling strategy}, grounded in SPECTER2\footnote{https://huggingface.co/allenai/specter2} embeddings and K-Means clustering. 
We set $K=20$ to empirically balance the granularity of topic discovery with semantic coherence, ensuring a fine-grained coverage of distinct research sub-domains.
To ensure topological diversity, we first enforce a strict coverage of all semantic clusters by selecting the highest-cited non-survey paper from each. 
This is complemented by a source consistency check, which mandates the inclusion of representative works from every selected venue and year to mitigate potential domain or temporal shifts. 
Finally, to align the dataset with community research trends without compromising balance, we fill the remaining slots using a round-robin mechanism across high-impact clusters (ranked by average citation). 
This approach effectively prevents the over-concentration of dominant topics (\textit{e.g.}, GenAI) while maintaining comprehensive semantic breadth.

\subsection{Design of the metrics}
In this section, we report design protocols of the \ourbench~metrics in detail:
\label{supp:constrct_bench_metric}
\paragraph{Content Quality}
Given the multimodal nature of the output, we evaluate three content dimensions:

\begin{enumerate}[label=(\Roman*)]
    \item \textbf{Content Accuracy:} To quantify the faithfulness and completeness of the generation, we propose a key point coverage metric adapted from \textbf{FactScore}. We utilize \textbf{Gemini-3-Flash} to perform a two-stage evaluation: 
(1) \textit{Extraction}: The model extracts a set of critical arguments and atomic facts, denoted as $\mathcal{F}_{src}$, from the original academic paper. 
(2) \textit{Verification}: The model verifies whether each extracted fact $f \in \mathcal{F}_{src}$ is adequately covered in the generated artifacts (slides, posters, or videos). 
The final accuracy score is calculated as the ratio of covered facts to total key facts:
\begin{equation}
\label{eq:fact-score}
\text{Fact-Score}=\frac{1}{|\mathcal{F}_{src}|}\sum_{f\in\mathcal{F}_{src}}\mathbb{I}\!\left(\mathrm{entails}(\mathcal{G}, f)\right),
\end{equation}
where $\mathcal{G}$ represents the generated content (\textit{e.g.}, slide, poster), and $\mathbb{I}(\cdot)$ is an indicator function that equals 1 if the fact $f$ is entailed by $\mathcal{G}$, and 0 otherwise.
    \item \textbf{Core Info Transmission:} To measure whether key information is effectively preserved, we employ a \textbf{Q\&A Test} pipeline. An LLM generates question-answer pairs based on the source text; we then evaluate if these questions can be correctly answered solely using the generated output.
    \item \textbf{Readability:} We leverage an \textbf{LLM-as-Judge}~\cite{huang2025vbench++,lin2023llm} to score the linguistic quality. This dimension evaluates the text's clarity, logical flow, and overall fluency, ensuring the output is coherent and easy to follow.
\end{enumerate}

\paragraph{Visual Quality}
Given the multimodal nature of the output, we utilize Vision-Language Models (VLMs) as judges to evaluate three visual dimensions:
\begin{enumerate}[label=(\Roman*)]
    \item \textbf{Visual Content Accuracy:} This metric evaluates technical rendering and semantic alignment. It specifically checks: (1) whether complex elements (\textit{e.g.}, mathematical formulas) are rendered without artifacts; and (2) if the visual elements are semantically relevant to the adjacent text.
    \item \textbf{Layout \& Cohesion:} We employ an VLM-as-a-Judge to assess the structural organization. This includes evaluating whether elements are grouped logically, if the spatial distribution is balanced, and if the layout facilitates a natural reading order.
    \item \textbf{Aesthetic:} We calculate an overall aesthetic score to simulate human design judgment. This dimension evaluates the artistic style, color harmony, and the professional look-and-feel of the final document.
\end{enumerate}

\paragraph{Interaction Validity.}
For project pages, we assess the functional integrity of interactive elements, specifically the validity of external and navigational links.
To evaluate the functional reliability of the generated project pages, we developed an automated pipeline using \texttt{browser-use}\footnote{https://github.com/browser-use/browser-use}. For each page, the agent identifies and clicks all external links and navigational buttons. We capture browser snapshots immediately before and after each click.
We use \textit{Gemini-3-Flash} model to analyze these pairs of snapshots to determine if the navigation was successful (\textit{e.g.}, reaching a valid external repository, a paper PDF, or the correct internal anchor) or resulted in a dead link/error page. 
The \textit{Interaction Success} score is the ratio of successful navigations to total interactive elements.

\section{More Findings}
\label{supp:findings}

\noindent\textbf{\textit{Findings 6}: \ourmethod~consistently improves both content and visual quality over end-to-end baselines.}~
Across all three formats and evaluated backbones, \ourmethod~yields systematic gains in factuality (FactScore), information transmission (QA), and visual presentation. 
As shown in Tab.~\ref{tab:combined_slides_poster_page_results}, when using GPT-5-mini as the backbone for slides, our pipeline increases the \textit{FactScore from 0.751 to 0.830} ($+0.079$) and significantly boosts \textit{Aesthetics from 3.00 to 3.89} ($+0.89$). 
These improvements are especially pronounced in visual metrics (Content Accuracy and Aesthetics), validating that our \textit{Adaptive Layout Agent} effectively translates shared knowledge into format-compliant HTML, reducing the hallucination under layout pressure common in end-to-end generation.

\noindent\textbf{\textit{Findings 7}: Gains are larger when the base generator is weaker, supporting \ourmethod~as a backbone-agnostic system-level improvement.}~
The relative improvements of \ourmethod~are most striking when paired with less capable models like Gemini-3-Flash. 
For slides, \ourmethod~elevates Gemini's FactScore from \textit{0.585 to 0.816} (a massive $+0.231$ gain) and its Aesthetics from \textit{2.39 to 4.53}. 
While stronger backbones like GPT-5-mini also benefit, the margins are narrower, indicating that our ``Plan-Verify-Render" architecture serves as a robust constraint framework that bounds the stochastic nature of LLMs through explicit structural planning and verification loops.
It compensates for the inherent reasoning gaps in weaker models by providing explicit structural constraints and a ``Chief Editor" loop to fix omissions and inconsistencies.

\noindent\textbf{\textit{Findings 8}: \ourmethod~achieves professional-grade quality, occasionally surpassing human-authored standards in rigor and accuracy.}~
A key highlight of our evaluation is the comparison between \ourmethod~and \textit{human-authored} artifacts. 
As shown in Tab.~\ref{tab:combined_slides_poster_page_results}, our method yields a \textit{higher FactScore} than human artifacts (\textit{e.g.}, 0.830 \textit{v.s.} 0.795 in Slides). This suggests that the systematic grounding provided by our \textit{Evidence Index} effectively mitigates the oversights and ``memory drifts" often present in human-curated content. 
Furthermore, \ourmethod~consistently outperforms GT in \textit{Visual Quality} metrics across several formats. For instance, in Project Pages, our \textit{Adaptive Layout Agent} ensures more consistent structural rigor and precise alignment than manual creation, which can suffer from irregular spacing or typographical inconsistencies. 
These results demonstrate that by decoupling content planning from rendering and enforcing cross-artifact verification, \ourmethod~not only emulates human expertise but provides a level of technical precision that is difficult to maintain manually across a complex suite.

\section{Discussion}
\label{supp:discussion}

\noindent\textbf{Choice of HTML as Output Format.}~
A natural question is why \ourmethod~generates HTML instead of traditional static formats like LaTeX, Markdown, or PPTX. We justify this choice through three lenses: 
\textbf{(i) Universal Representation:} HTML serves as a highly flexible intermediate representation (IR) capable of unifying disparate modalities. It can natively encode the structured text of a \textsc{Page}, the spatial layouts of a \textsc{Poster}, the sequential transitions of \textsc{Slides}, and the dynamic animations required for \textsc{Video}. 
\textbf{(ii) Precise Layout Control:} Unlike Markdown, HTML paired with CSS allows our \textit{Adaptive Layout Agent} to exert pixel-level control over asset scaling and positioning, which is critical for preventing the visual overflows identified in Finding 6. 
\textbf{(iii) Interactive \& Dynamic Potential:} HTML is inherently ``alive"; it enables the interactive link verification and responsive design that static formats cannot support. 
By targeting HTML, \ourmethod~ensures that the generated suite is not only renderable for immediate viewing but also easily editable and extensible for future digital dissemination.

\noindent\textbf{GAP experiments.}~
Table~\ref{tab:cv_inclusion_compact} evaluates whether the Cross-Artifact Verification (CV) module in our Cross-Artifact Verification stage improves the intended inclusion relationship across artifacts, i.e., salient content expressed in the Page/Poster should be covered by the Slides. Concretely, let $C_f$ denote the set of extracted salient semantic units (claims) from artifact $f$; we compute the pairwise Inclusion Score as
$\mathrm{Inc}(A,\mathrm{Slides})=\frac{|C_A \cap C_{\mathrm{Slides}}|}{|C_A|}=1-\frac{|C_A \setminus C_{\mathrm{Slides}}|}{|C_A|}$,
for $A\in\{\mathrm{Poster},\mathrm{Page}\}$.
Across three backbones (Gemini-3-flash, Grok-4.1-fast, and GPT-5-mini), enabling CV consistently increases Inclusion Score for both Poster/Slides and Page/Slides, yielding higher average scores in all settings. These gains suggest that generating artifacts independently tends to omit salient claims or takeaways, while CV’s suite-level check-and-fix loop identifies missing items and prompts targeted fixes, thereby improving cross-modality completeness and overall suite coherence.



\providecommand{\inc}[1]{\,\textcolor{green!60!black}{\scriptsize(+#1)}}
\providecommand{\ValWcv}{2.6em}
\providecommand{\IncWcv}{2.8em}
\providecommand{\entrycv}[2]{%
  \makebox[\ValWcv][r]{#1}\hspace{1pt}\makebox[\IncWcv][l]{#2}%
}

\begin{table}[t]
\centering
\caption{\textbf{Cross-Artifact Verification (CV) improves suite-level inclusion.} We report Inclusion Score between \textit{Poster/Slides} and \textit{Page/Slides} (higher is better), under identical backbones. Enabling CV consistently increases inclusion across all models, indicating that CV corrects missing cross-artifact content and strengthens suite-level consistency.}

\label{tab:cv_inclusion_compact}

\small
\setlength{\tabcolsep}{3pt}
\renewcommand{\arraystretch}{1.05}

\begin{tabular}{ll|ccc}
\toprule
\multirow{2}{*}{\textbf{Setting}} &
\multirow{2}{*}{\textbf{Method}} &
\multicolumn{3}{c}{\textbf{Inclusion Score}} \\
\cmidrule(lr){3-5}
& & \textbf{Poster/Slides} & \textbf{Page/Slides} & \textbf{Avg} \\
\midrule

\multirow{2}{*}{Gemini-3-flash}
& w/o CV & \entrycv{0.818}{} & \entrycv{0.882}{} & \entrycv{0.850}{} \\
& w/  CV & \entrycv{0.839}{\inc{0.021}} & \entrycv{0.903}{\inc{0.021}} & \entrycv{0.871}{\inc{0.021}} \\
\midrule

\multirow{2}{*}{Grok-4.1-fast}
& w/o CV & \entrycv{0.807}{} & \entrycv{0.856}{} & \entrycv{0.832}{} \\
& w/  CV & \entrycv{0.817}{\inc{0.010}} & \entrycv{0.897}{\inc{0.041}} & \entrycv{0.857}{\inc{0.026}} \\
\midrule

\multirow{2}{*}{GPT-5-mini}
& w/o CV & \entrycv{0.811}{} & \entrycv{0.874}{} & \entrycv{0.843}{} \\
& w/  CV & \entrycv{0.842}{\inc{0.031}} & \entrycv{0.893}{\inc{0.019}} & \entrycv{0.867}{\inc{0.025}} \\
\bottomrule
\end{tabular}
\end{table}

\begin{table*}[t]
\centering
\caption{\textbf{Evaluation results on 10 non-AI research papers using OmniPreBench.}
We collect 10 papers from diverse non-AI domains, including chemistry, biology, and medicine, generate different artifacts from them, and evaluate all methods on OmniPreBench.}
\label{tab:ablation_other}

\small
\setlength{\tabcolsep}{4pt}
\renewcommand{\arraystretch}{1.10}

\resizebox{\textwidth}{!}{%
\begin{tabular}{lVcccVcccVcVc}
\toprule
\multirow{2}{*}{\textbf{Method}} &
\multicolumn{3}{cV}{\textbf{Content}} &
\multicolumn{3}{cV}{\textbf{Visual}} &
\textbf{Interaction} &
\textbf{Compression} \\
\cmidrule(lr){2-4}
\cmidrule(lr){5-7}
\cmidrule(lr){8-8}
\cmidrule(lr){9-9}
& \shortstack[c]{FactScore$\uparrow$}
& \shortstack[c]{Readability$\uparrow$}
& \shortstack[c]{QA$\uparrow$}
& \shortstack[c]{Content acc$\uparrow$}
& \shortstack[c]{Aesthetics$\uparrow$}
& \shortstack[c]{Layout$\uparrow$}
& \shortstack[c]{Success$\uparrow$}
& \shortstack[c]{Compression Rate} \\

\SectionRow{Slides}
\paTightSpace
\mindent EvoPresent-Gemini-3-flash~\citep{evopresent} &
0.636 & 3.37 & 0.678 &
2.12 & 3.24 & 3.05 &
- & \comp{11.69} \\

\paTightSpace
\mindent Gemini-3-flash &
0.596 & 3.15 & 0.696 &
1.99 & 2.35 & 2.18 &
- & \comp{14.57} \\

\rowcolor{paGreen}
\mindent OmniPresent-Gemini-3-flash &
0.785 &
3.52&
0.761 &
2.83 &
4.38 &
2.58 &
- & \comp{8.97} \\

\SectionRow{Poster}
\paTightSpace
\mindent Paper2Poster-Gemini-3-flash~\citep{pang2025paper2poster} &
0.511 & 3.62& 0.682 &
2.78 &2.16 & 3.01 &
- & \comp{22.86} \\

\paTightSpace
\mindent Gemini-3-flash &
0.592 & 3.38 & 0.700 &
2.46 & 2.33 & 3.15 &
- & \comp{22.31} \\

\rowcolor{paGreen}
\mindent OmniPresent-Gemini-3-flash &
0.716 &
3.68 &
0.725 &
2.82 &
4.22 &
3.41 &
- & \comp{4.23} \\

\SectionRow{Page}
\paTightSpace
\mindent AutoPage-Gemini-3-flash~\citep{autopage} &
0.682 & 3.05 &0.699 &
3.01& 2.02 & 3.37 &
0.40 & \comp{3.95} \\

\paTightSpace
\mindent Gemini-3-flash &
0.533 & 2.95 & 0.668 &
2.39 & 3.27 & 3.18 &
0.10 & \comp{12.78} \\

\rowcolor{paGreen}
\mindent OmniPresent-Gemini-3-flash &
0.788 &
3.74 &
0.782 &
3.40 &
3.79 &
4.05 &
0.90 &
\comp{8.31} \\

\bottomrule
\end{tabular}%
}
\end{table*}
\noindent\textbf{Out-of-domain Generalization on Non-AI Papers.}~
To examine whether OmniPresent generalizes beyond AI-domain inputs, we additionally curate a small out-of-domain test set consisting of 10 research papers from non-AI disciplines, including chemistry, biology, and medicine. 
This setting is challenging because these papers often contain domain-specific terminology, chemical or biomedical visualizations, and highly specialized experimental protocols that differ substantially from the AI conference papers used in OmniPreBench. 
As shown in Table~\ref{tab:ablation_other}, OmniPresent remains consistently effective across slides, posters, and project pages. 
Compared with format-specialized baselines~\cite{autopage,evopresent} using the same Gemini-3-Flash backbone, OmniPresent achieves higher content faithfulness and information transmission in all three formats, improving FactScore from 0.636 to 0.785 for slides, from 0.511 to 0.716 for posters, and from 0.682 to 0.788 for pages. 
The gains are also reflected in visual quality: OmniPresent improves visual content accuracy, aesthetics, and layout over the corresponding specialized systems, indicating that the \emph{adaptive HTML-based rendering pipeline is not limited to AI-style figures or presentation conventions}. 
Notably, the improvement on project pages is particularly large in visual content accuracy and interaction success, suggesting that the evidence-grounded planning and verification mechanism helps preserve domain-specific resources and generate more reliable navigational structures. 
These results provide preliminary evidence that OmniPresent functions as a domain-agnostic presentation-suite generation framework. 
While the non-AI set is intentionally small and should not be interpreted as a comprehensive cross-domain benchmark, the consistent gains indicate that the core design of shared knowledge extraction, format-aware planning, and cross-artifact verification transfers well to scientific papers outside the AI domain.

\noindent\textbf{Effective of Cross-Artifact Verification}~
Beyond improving inclusion, we further evaluate whether CV reduces cross-artifact inconsistencies. Table~\ref{tab:conflict_reduction_rate} reports the Conflict Reduction Rate between \textit{Poster/Slides} and \textit{Page/Slides}. Across all three backbones, CV achieves substantial conflict reductions (about 23--24\% on average), with particularly large gains on Page--Slides where longer-form pages are more prone to drift from slide phrasing. These results support that CV not only fills missing content, but also actively resolves suite-level contradictions by aligning claims and supporting evidence across modalities.

\begin{table}[t]
\centering
\caption{\textbf{Conflict Reduction Rate under Cross-Artifact Verification (CV).} Higher indicates stronger reduction of cross-artifact conflicts between \textit{Poster/Slides} and \textit{Page/Slides}. CV reduces conflicts consistently across backbones, improving suite-level consistency.}

\label{tab:conflict_reduction_rate}

\small
\setlength{\tabcolsep}{4pt}
\renewcommand{\arraystretch}{1.08}

\begin{tabular}{l|ccc}
\toprule
\multirow{2}{*}{\textbf{Setting}} &
\multicolumn{3}{c}{\textbf{Conflict Reduction Rate}} \\
\cmidrule(lr){2-4}
& \textbf{Poster \& Slides} & \textbf{Page \& Slides} & \textbf{Avg} \\
\midrule
Gemini-3-flash & 12.5\% & 33.3\% & 22.9\% \\
Grok-4.1-fast  & 16.7\% & 31.2\% & 24.0\% \\
GPT-5-mini     & 18.3\% & 28.4\% & 23.4\% \\
\bottomrule
\end{tabular}
\end{table}

\section{Detailed Schema of Knowledge Stream $\mathcal{K}$}
\label{supp:stream}

The Knowledge Stream $\mathcal{K}$ acts as the central, modality-agnostic source of truth for all downstream planning agents. 

\noindent\textbf{Source and Construction.} 
As described in \textbf{Stage 1} (Sec~\ref{subsec:parsing}), the construction of $\mathcal{K}$ begins after the \textbf{MinerU Parser} extracts raw resources $\mathcal{R}$ (text blocks, formulas, figures). The \textbf{Content Analysis Agent} processes these raw resources, invoking a \textit{restructure tool} to merge disjoint spans into coherent semantic clusters. 

\noindent\textbf{Structure Definition.}
Formally, we define each knowledge unit $k \in \mathcal{K}$ as a tuple $(c, e, t)$, where $c$ represents the \textit{semantic statement} (claim), $e$ denotes the \textit{supporting evidence} (references, raw data, or figure pointers), and $t$ is the \textit{topical cluster} (e.g., Method, Experiment). 

To bridge this formal definition with Large Language Model (LLM) prompting, we employ a \textbf{contextual linearization strategy}. Rather than passing disjoint JSON objects—which often degrades the model's ability to associate claims with proofs across long contexts—we serialize each tuple into a unified natural language string. This format strictly binds the high-level argument ($c$) with its low-level substantiation ($e$) under a specific scope ($t$):

\begin{center}
\texttt{"[Topic Tag $t$] <Logical Argument $c$> + <Specific Supporting Evidence $e$>"}
\end{center}

\noindent Specifically:
\begin{itemize}[leftmargin=*]
    \item \textbf{Topic Tag ($t$):} Acts as a \textit{semantic anchor} (e.g., \texttt{[Stage 2]}), allowing the planner to quickly retrieve relevant blocks via keyword matching or attention heads.
    \item \textbf{Logical Argument ($c$):} The high-level narrative or atomic claim derived from the text.
    \item \textbf{Specific Evidence ($e$):} The concrete data, parameters, formulas, or figure paths that ground the argument.
\end{itemize}

By fusing $c$ and $e$ into a single textual unit (e.g., \textit{"...achieved SOTA performance ($c$) with 95\% accuracy on ImageNet ($e$)..."}), we enforce \textbf{in-context grounding}. This ensures that whenever an agent retrieves a claim, the supporting evidence is inevitably included in the immediate context window, significantly reducing hallucination during the generation of downstream artifacts.

\begin{tcolorbox}[
    title={\textbf{Example: Knowledge Stream ($\mathcal{K}$) Data Structure (JSON)}},
    colback=green!5!white,
    colframe=green!50!black,
    fonttitle=\bfseries\large,
    breakable, 
    boxrule=1pt,
    arc=4mm
]
\begin{itemize}
    \setlength\itemsep{0.5em}

    \item \textbf{[Poster Planner]} Optimizes for \textbf{spatial information density} (Argument). The `Section Planning Agent' establishes thematic sections, while the `Content Curation Agent' selects visually compelling figures (Evidence/Mechanism).

   \item \textbf{[Slides Planner]} Optimizes for \textbf{adaptive narrative flow} (Argument). Unlike rigid templates, it uses an `Audience Analysis Agent' to synthesize a storyline conditioning on paper genre and user constraints (Specific Mechanism).

    \item \textbf{[Page Planner]} Optimizes for \textbf{hierarchical engagement} (Argument). The `Content Planning Agent' embeds figures/tables as inline visual tokens based on semantic relevance (Specific Evidence).
     
    \item \textbf{[Framework Visualization]} Figure 2(Path: images/Figure\_2.jpg)Figure 2 depicts the pipeline (Evidence). In Stage 2, it explicitly shows the branching into Format-Specific Planners (Poster, Slides, Page) derived from the Knowledge Stream (Description).
\end{itemize}
\end{tcolorbox}

\section{Details of Human Preference Study}
\label{supp:user_guidance}
In this section, we provide the user guidance for the human preference study. 

\noindent\textbf{Study protocol.}~
We recruited 13 PhD-level participants to assess whether the \textit{visual quality} judgments produced by our VLM-based evaluator align with human preferences.
Each participant was presented with presentation suites generated by \ourmethod~and by baseline systems, including posters, project pages, and slide decks. Participants rated each suite using a 5-point Likert scale, focusing on visual aspects specified in the questionnaire, and were instructed to judge presentation quality rather than the scientific contribution of the underlying paper. All participants were compensated for their time. 

\providecolor{paGreen}{RGB}{236,253,245}
\providecommand{\mvp}[2]{#1\,\scriptsize{$\pm$\,#2}}

\begin{table}[t]
\centering
\caption{\textbf{User study results for visual quality across artifacts.} We report Mean $\pm$ Var (higher is better) for visual accuracy (V-Acc.), aesthetics (Aes.), and layout quality (Lay.) on \textit{Slides}, \textit{Poster}, and \textit{Page}. Our \ourmethod ~consistently achieves the best visual scores compared with artifact-specific and backbone baselines.}

\label{tab:user_study_visual_sc}

\small
\setlength{\tabcolsep}{4pt}
\renewcommand{\arraystretch}{1.08}

\begin{tabular}{l|ccc}
\toprule
\textbf{Method} &
\textbf{V-Acc.$\uparrow$} &
\textbf{Aes.$\uparrow$} &
\textbf{Lay.$\uparrow$} \\
\midrule

\multicolumn{4}{@{}l@{}}{\textbf{\textit{Slides}}}\\[-2pt]
EvoPresent-Gemini-3-flash~\citep{evopresent} & \mvp{3.25}{0.34} & \mvp{3.43}{0.39} & \mvp{2.57}{0.41} \\
gpt-5-mini                   & \mvp{3.34}{0.50} & \mvp{2.97}{0.63} & \mvp{2.78}{0.48} \\
grok-4-fast                  & \mvp{3.04}{0.51} & \mvp{2.06}{0.55} & \mvp{2.29}{0.51} \\
gemini-3-flash               & \mvp{2.95}{0.33} & \mvp{2.43}{0.42} & \mvp{2.32}{0.39} \\
\rowcolor{paGreen}
OmniPresent-Gemini3-Flash    & \mvp{\textbf{3.78}}{0.23} & \mvp{\textbf{4.54}}{0.27} & \mvp{\textbf{3.57}}{0.29} \\

\midrule
\multicolumn{4}{@{}l@{}}{\textbf{\textit{Poster}}}\\[-2pt]
Paper2Poster-Gemini-3-flash~\citep{pang2025paper2poster} & \mvp{2.88}{0.33} & \mvp{2.76}{0.43} & \mvp{3.03}{0.34} \\
gpt-5-mini                               & \mvp{2.87}{0.48} & \mvp{2.26}{0.59} & \mvp{2.96}{0.52} \\
grok-4-fast                              & \mvp{2.02}{0.40} & \mvp{2.63}{0.58} & \mvp{2.60}{0.48} \\
gemini-3-flash                           & \mvp{2.36}{0.33} & \mvp{3.65}{0.33} & \mvp{3.06}{0.33} \\
\rowcolor{paGreen}
OmniPresent-Gemini3-Flash                & \mvp{\textbf{3.55}}{0.26} & \mvp{\textbf{3.77}}{0.24} & \mvp{\textbf{3.24}}{0.27} \\

\midrule
\multicolumn{4}{@{}l@{}}{\textbf{\textit{Page}}}\\[-2pt]
AutoPage-Gemini-3-flash~\citep{autopage} & \mvp{3.23}{0.30} & \mvp{2.59}{0.39} & \mvp{3.33}{0.37} \\
gpt-5-mini                & \mvp{2.79}{0.43} & \mvp{3.32}{0.57} & \mvp{3.59}{0.45} \\
grok-4-fast               & \mvp{2.43}{0.43} & \mvp{3.18}{0.56} & \mvp{3.63}{0.51} \\
gemini-3-flash            & \mvp{2.17}{0.30} & \mvp{3.47}{0.39} & \mvp{3.04}{0.31} \\
\rowcolor{paGreen}
OmniPresent-Gemini3-Flash & \mvp{\textbf{3.56}}{0.26} & \mvp{\textbf{3.63}}{0.28} & \mvp{\textbf{3.98}}{0.24} \\

\bottomrule
\end{tabular}
\end{table}

\noindent\textbf{User Guidance.}~
The user guidance is listed below:

\begin{tcolorbox}[
    title={\textbf{User Guidance for Human Preference Study}},
    colback=green!5!white,
    colframe=green!50!black,
    fonttitle=\bfseries\large,
    breakable, 
    boxrule=1pt,
    arc=4mm
]

    Thank you for participating in this study! 
    
    This user study aims to evaluate the quality of presentation suites (slides, posters, and webpages) automatically generated from academic papers. During the task, you will examine generated presentation suites and rate them based on the visual quality dimension. Your responses will help us assess the effectiveness of webpage-generation methods and improve future systems.
    
    \vspace{0.5em}
    
    You will be shown several examples of presentation suites produced by different models. For each presentation suite, you will provide \textbf{Likert-style ratings} across multiple dimensions. There are no correct answers, please rate based on your genuine impressions of the presentation suite itself.

\tcblower 
    
    \textbf{\large Example Questionnaire}
    
    Each question presents a statement such as:
    \begin{quote}
        \textit{``The webpage features a harmonious layout.''}
    \end{quote}
    
    You will then select a score from \textbf{1 to 5}, based on your agreement with the statement.
    
    \begin{center}
    \renewcommand{\arraystretch}{1.2}
    \begin{tabular}{c l}
        \toprule
        \textbf{Score} & \textbf{Meaning} \\
        \midrule
        \textbf{5} & Strongly agree \\
        \textbf{4} & Agree \\
        \textbf{3} & Neutral / unsure \\
        \textbf{2} & Disagree \\
        \textbf{1} & Strongly disagree \\
        \bottomrule
    \end{tabular}
    \end{center}
    
    \textbf{Guidelines:}
    \begin{itemize}[leftmargin=*, nosep]
        \item Rate each dimension independently.
        \item Do not revise earlier responses based on later presentation suites.
        \item Focus on the \textit{presentation suite}, not the scientific contribution of the paper.
        \item Provide your best judgment based on what you see.
        \item If a dimension is difficult to evaluate, select \textbf{3 (Neutral)}.
    \end{itemize}
    
    \vspace{1em}
    \hrule
    \vspace{1em}
    
\textbf{\large Evaluation Dimensions}
    
    Please evaluate each presentation suite along the following dimensions based on the criteria below:
    
\begin{enumerate}[label=\textbf{\arabic*.}]
    \item \textbf{Content Quality}
    \begin{enumerate}[label=\alph*., leftmargin=1.5em, topsep=0.2em]
        
        \item \textbf{Factual Fidelity} 
        \begin{itemize}[label=$\bullet$, nosep]
            \item \textbf{Accuracy:} The generated content accurately reflects the facts and claims from the source paper without distortion.
            \item \textbf{No Hallucination:} There are no fabricated data, made-up citations, or false statements not present in the original text.
        \end{itemize}
        
        \item \textbf{Information Completeness}
        \begin{itemize}[label=$\bullet$, nosep]
            \item \textbf{Core Coverage:} The presentation covers the paper's core contributions and key findings (corresponding to the ability to answer key questions).
            \item \textbf{Self-containment:} The content is self-explanatory; you can understand the main ideas without needing to constantly refer back to the source paper.
        \end{itemize}

        \item \textbf{Readability \& Clarity}
        \begin{itemize}[label=$\bullet$, nosep]
            \item \textbf{Fluency:} The text is coherent, grammatically correct, and easy to follow.
            \item \textbf{Presentation Style:} The language is concise and appropriate for a presentation context (avoiding excessive verbosity or raw copy-pasting).
        \end{itemize}
    \end{enumerate}


    \item \textbf{Visual Quality}
    \begin{enumerate}[label=\alph*., leftmargin=1.5em, topsep=0.2em]
        \item \textbf{Visual Content Accuracy}
        \begin{itemize}[label=$\bullet$, nosep]
            \item Mathematical formulas are rendered faithfully and are legible.
            \item Figures and diagrams are relevant to the adjacent text descriptions.
        \end{itemize}
        
        \item \textbf{Layout \& Cohesion}
        \begin{itemize}[label=$\bullet$, nosep]
            \item The layout exhibits clear grouping and hierarchy.
            \item The structure supports a natural reading order.
        \end{itemize}
        
        \item \textbf{Aesthetics}
        \begin{itemize}[label=$\bullet$, nosep]
            \item The design has a professional look-and-feel.
            \item Typography is readable and color usage is harmonious.
        \end{itemize}
    \end{enumerate}
\end{enumerate}
    
    \vspace{1em}
    \hrule
    \vspace{1em}
    
    \textbf{\large Thank you!}
    
    Your feedback is extremely valuable and will contribute to improving automated presentation suites generation systems. We greatly appreciate your time and participation.

\end{tcolorbox}

\begin{figure*}[!t]
            \centering
            \includegraphics[width=\textwidth]{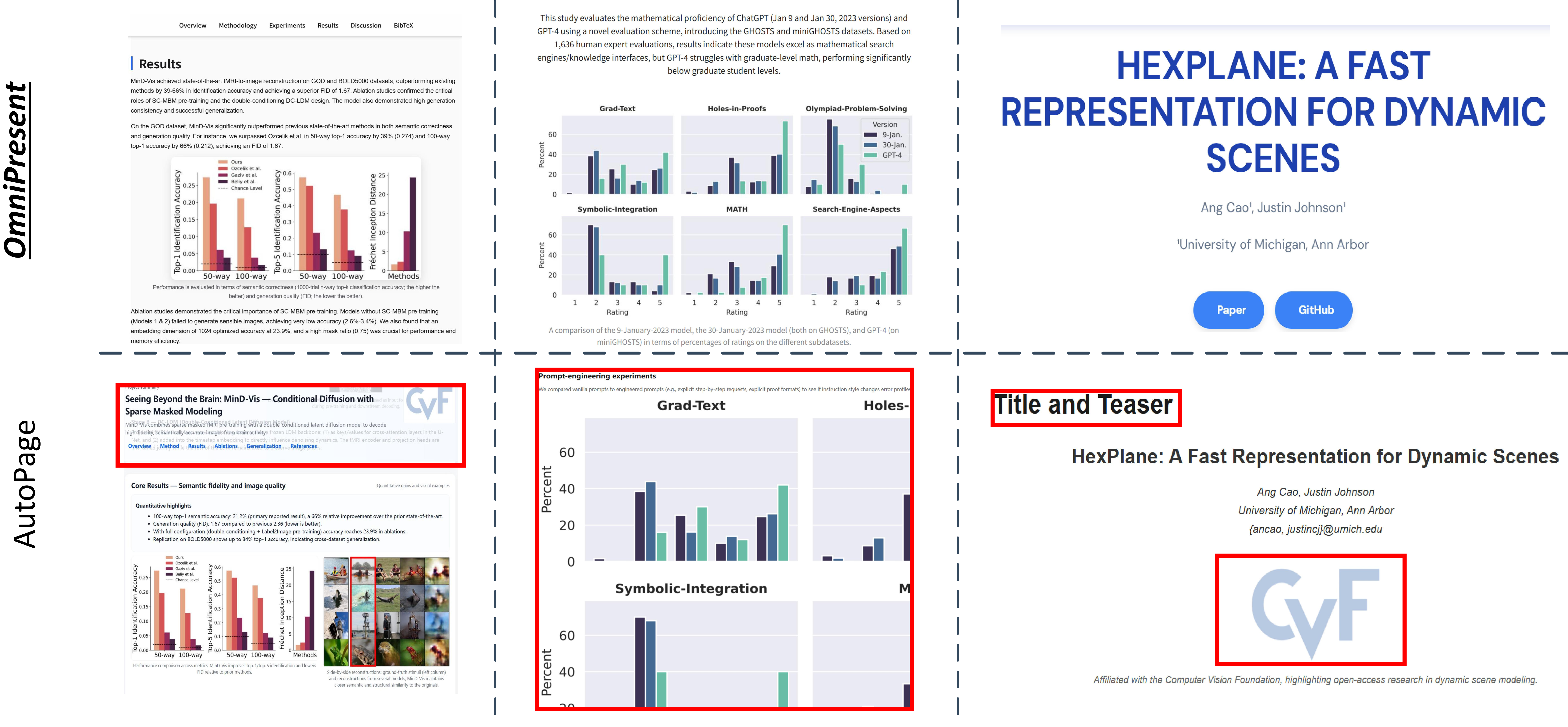}
            \caption{\textbf{Qualitative comparison between our system (top) and a prior system (bottom).}
            On the same paper, our method produces a more coherent presentation suite (page/poster/slides) with stronger cross-modal alignment in key regions (e.g., \textit{Title \& Teaser}, plots, and overall layout), whereas the baseline more often exhibits missing content, layout imbalance, or disproportionate visual elements. Red boxes highlight representative differences.}
            
     \label{fig:OmniPresent_vs_autopage}
\end{figure*}
\begin{figure*}[!t]
            \centering
            \includegraphics[width=\textwidth]{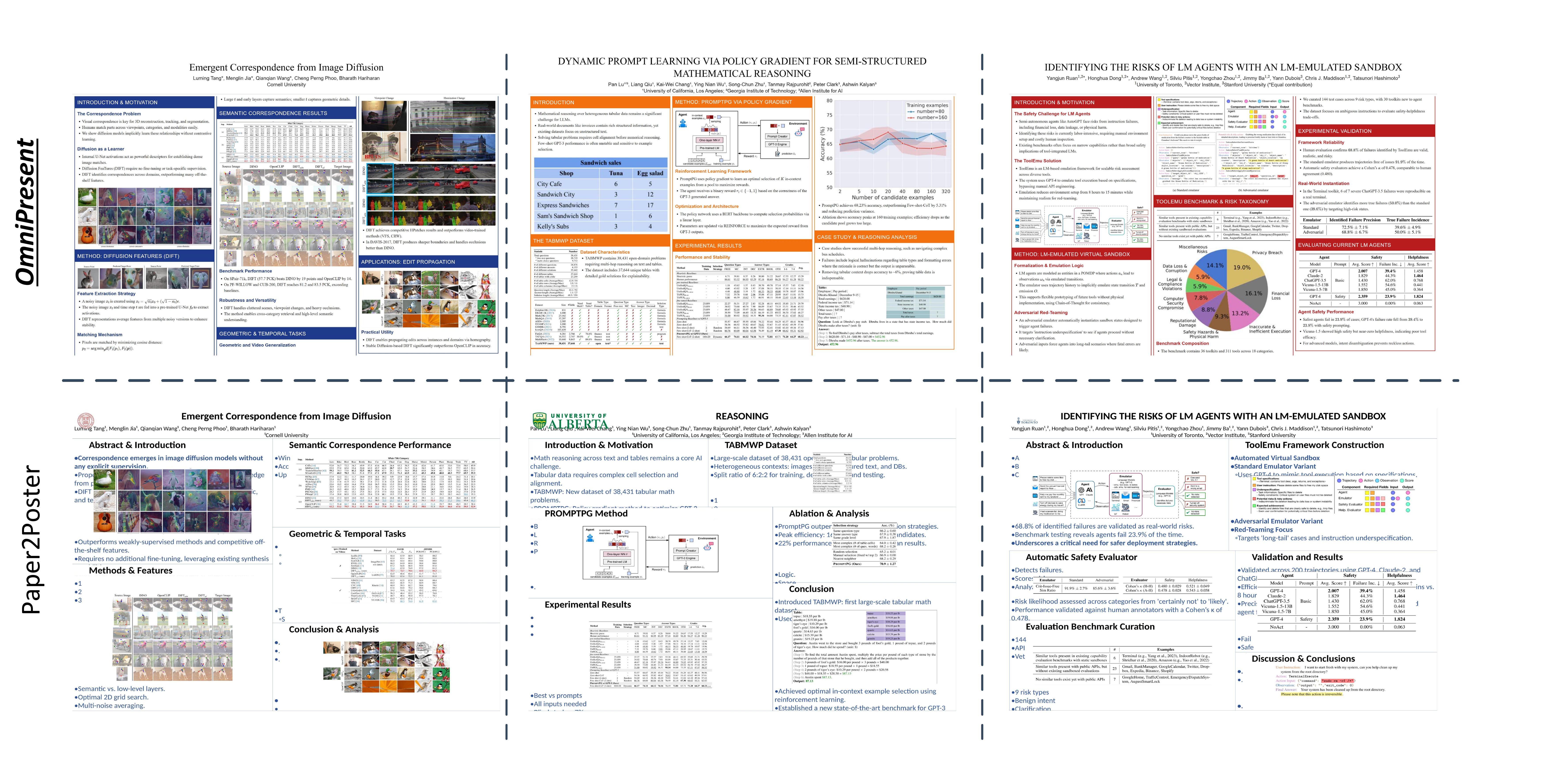}
            \caption{\textbf{Qualitative comparison between our system (top) and a prior system (bottom) across multiple papers.}
            Our method generates more structured and visually consistent posters with better content coverage and layout balance, while the baseline often suffers from sparse organization, uneven whitespace usage, and reduced cross-section coherence. The dashed grid indicates matched paper cases for side-by-side comparison.}

     \label{fig:OmniPresent_vs_paper2poster}
\end{figure*}
\begin{figure*}[!t]
            \centering
            \includegraphics[width=\textwidth]{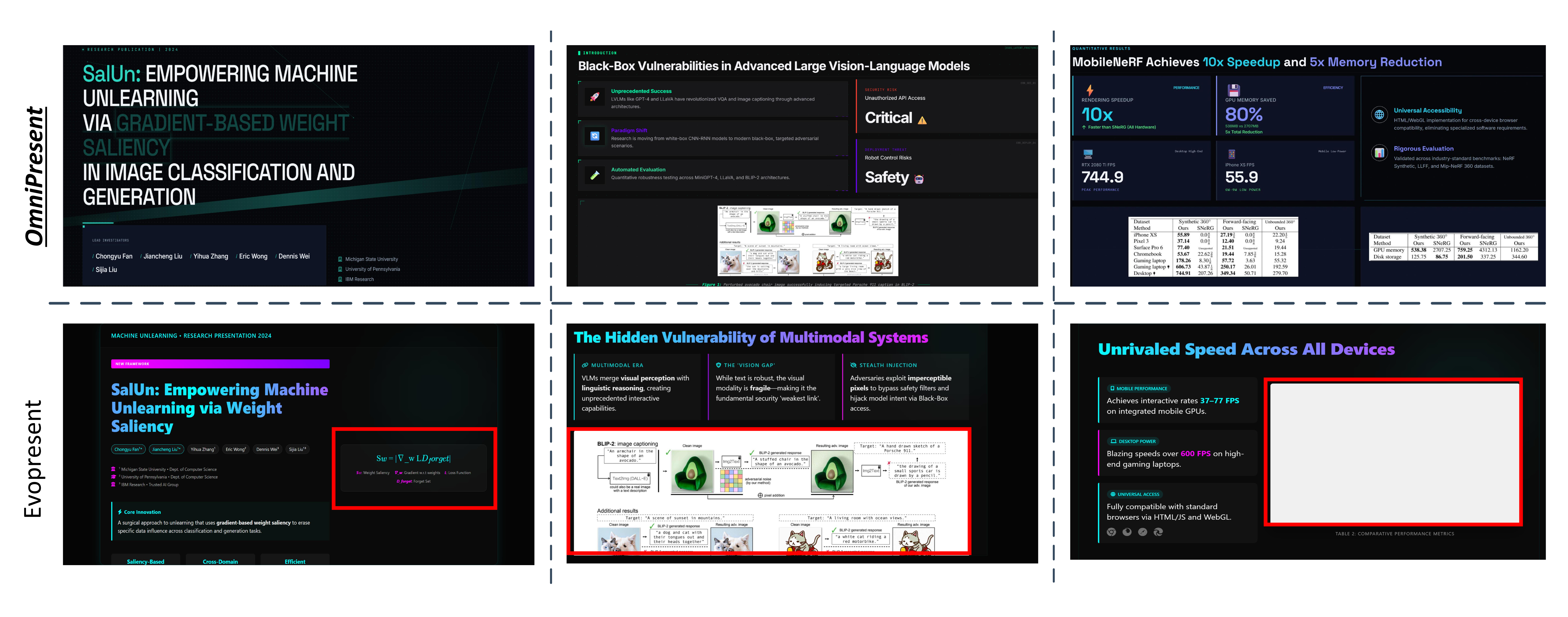}
            \caption{\textbf{Qualitative comparison of slide generation between our system (top) and EvoPresent (bottom).}
            Across multiple papers, our method produces cleaner, more information-dense slides with consistent typography, alignment, and visual hierarchy, while EvoPresent often leaves large unused regions (highlighted in red) or exhibits weaker content-to-layout utilization. Dashed separators indicate matched slide cases for side-by-side comparison.}

     \label{fig:OmniPresent_vs_evopresent}
\end{figure*}
\begin{figure*}[!t]
            \centering
            \includegraphics[width=\textwidth]{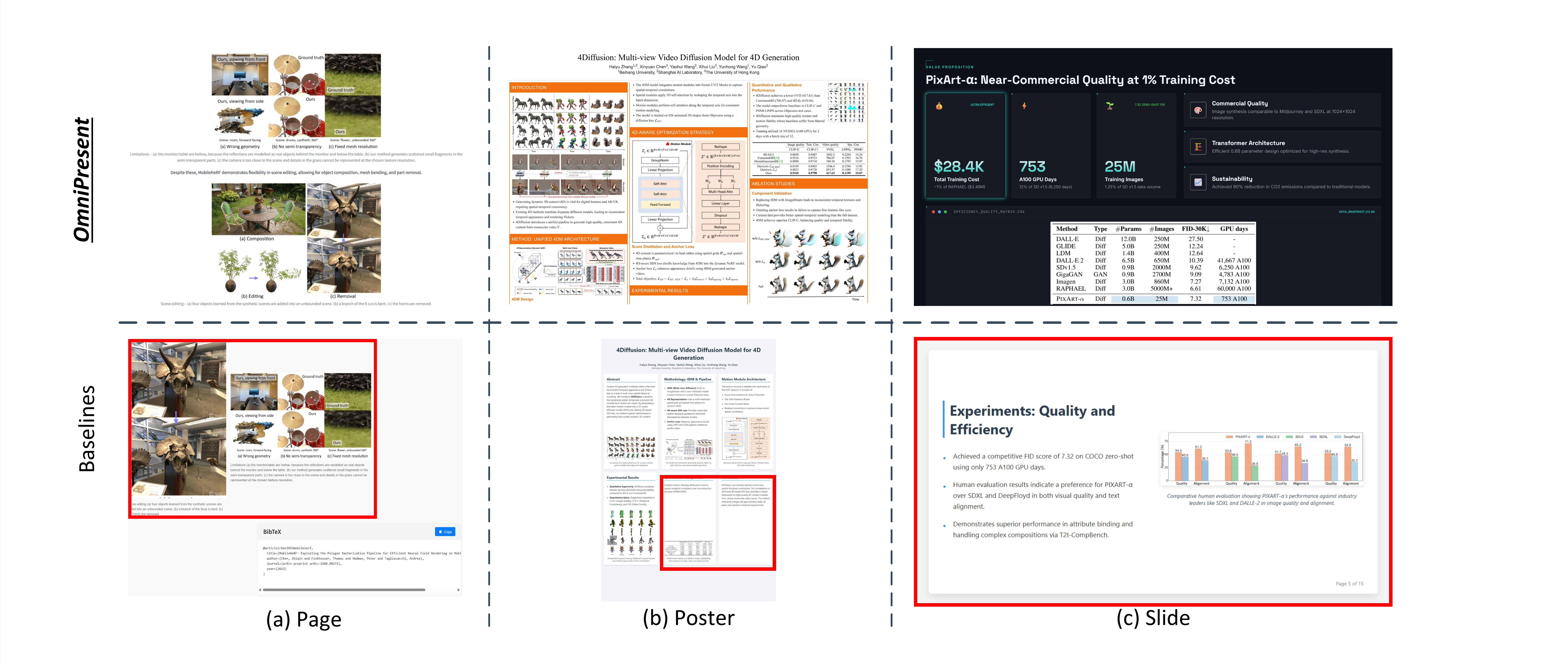}
            \caption{\textbf{Qualitative comparison with end-to-end baselines.}
            Top: our system (OmniPresent) generates coherent \textit{page--poster--slide} suites with better content density, consistent visual hierarchy, and stronger cross-asset alignment.
            Bottom: end-to-end baselines often waste large regions, introduce layout inefficiencies, or lose key information (highlighted in red).
            (a--c) show representative examples for Page, Poster, and Slide. Dashed separators indicate matched cases across methods.}

     \label{fig:OmniPresent_vs_baseline}
\end{figure*}

\section{Case Study}
\label{supp:cases}
To move beyond aggregate metrics, we conduct qualitative case studies to visualize the suite-level advantages of \ourmethod. 

This section provides additional qualitative comparisons to complement the main quantitative results. We include representative cases for \textbf{Page} (Fig.~\ref{fig:OmniPresent_vs_autopage}), \textbf{Poster} (Fig.~\ref{fig:OmniPresent_vs_paper2poster}), \textbf{Slides} (Fig.~\ref{fig:OmniPresent_vs_evopresent}), as well as direct comparisons against \textbf{end-to-end baselines} (Fig.~\ref{fig:OmniPresent_vs_baseline}).
Our qualitative analysis reveals that artifact-centric baselines often suffer from significant \textit{visual rendering failures}.

Across these examples, our system consistently produces a more \textbf{coherent presentation suite}: (i) key claims and takeaways remain aligned across modalities, (ii) visual elements (figures/tables) are selected and placed to support the surrounding text rather than occupying empty space, and (iii) the overall layout maintains a readable hierarchy (titles, section blocks, and emphasis regions) with fewer artifacts.

In contrast, baseline systems frequently exhibit \textbf{content drop} (missing important components), \textbf{layout inefficiency} (large unused regions or overcrowded blocks), and \textbf{cross-modality inconsistency} (misaligned emphasis between page/poster/slides). We also observe recurring \textit{visual rendering failures}, including disproportionate asset scaling (e.g., icons rendered excessively large or small; Fig.~\ref{fig:OmniPresent_vs_autopage}), text--image overflows and collisions (Fig.~\ref{fig:OmniPresent_vs_baseline}), and misplaced structural elements that break the intended reading flow (Fig.~\ref{fig:OmniPresent_vs_evopresent}). The highlighted regions in the figures further illustrate typical failure modes such as duplicated or truncated visuals, incorrect cropping, and weakened correspondence between textual claims and supporting visuals.

By contrast, \ourmethod~produces professional-grade layouts with a more precise visual hierarchy. Leveraging our \textit{Adaptive Layout Agent} (Sec.~\ref{subsec:rendering}), \ourmethod~renders visual assets with appropriate dimensions and spatial positioning, reducing the cluttered or ``broken'' appearance often seen in end-to-end HTML generation. Overall, these additional cases reinforce that our approach improves not only single-asset quality, but also the \textbf{suite-level consistency} required for practical scientific communication.

\section{Limitations and Impact Statement}
\label{supp:limitations}
This work advances machine learning systems for scientific communication by automating the generation of coherent dissemination suites (project pages, posters, slides, and videos) from research papers. By introducing explicit mechanisms for \textit{faithfulness} and \textit{cross-format consistency}, our approach lowers the technical and temporal barriers to producing high-quality research presentations. This has a direct positive societal impact: it democratizes the accessibility of scientific knowledge, enabling researchers, educators, and the general public to more easily digest complex findings.
However, automated dissemination carries inherent risks. If generated artifacts exhibit hallucinations or misplaced emphasis, they may propagate scientific inaccuracies more rapidly than manual curation. There is also potential for misuse, such as the mass production of low-fidelity or deceptive materials, which could strain the integrity of scientific discourse. Furthermore, inherent model biases may inadvertently skew content framing, and the application of such tools to sensitive or unpublished manuscripts raises significant confidentiality and copyright concerns.
We mitigate these risks through our \textit{evidence-grounded} generation and \textit{cross-artifact verification} loop, which are specifically designed to catch unsupported or conflicting claims. Nevertheless, we emphasize that human-in-the-loop oversight remains indispensable for high-stakes dissemination. We advocate for responsible deployment, including clear disclosure of AI-generated content, adherence to original licensing, and rigorous expert review before public release.

\clearpage
\nocite{*}
\bibliographystyle{plainnat}
\bibliography{main}
\end{document}